\documentclass[11pt,letterpaper,onecolumn]{article}
\usepackage[margin=0.85in]{geometry}
\usepackage[numbers,sort&compress]{natbib}

\usepackage{microtype}
\providecommand{\onecolumngrid}{}
\providecommand{\affiliation}[1]{}

\usepackage{amsmath}
\usepackage{amssymb}
\usepackage{array}
\usepackage{booktabs}
 \usepackage{xcolor}

\usepackage{bbold}
\usepackage{xcolor}
\usepackage{dcolumn}
\usepackage{tikz}
\usepackage{bm}
\usepackage{pifont}
\usepackage{pgfplots}

\usepackage{setspace}
\usepackage{soul}
\pgfplotsset{compat=1.5}
\usepackage{graphicx}

\definecolor{dodgerblue}{HTML}{1E90FF}
\definecolor{columbiablue}{HTML}{87AFC7}
 
\usepackage[colorlinks=true,citecolor=dodgerblue,linkcolor=dodgerblue,urlcolor=dodgerblue]{hyperref}
\usepackage{cleveref}
\usepackage{blindtext}
\usepackage{mathtools}
\usepackage{titletoc}

\newcommand{\CsI}{CsI$_3$}

\DeclarePairedDelimiterX\braket[2]{\langle}{\rangle}{#1\,\delimsize\vert\,\mathopen{}#2}

\begin{document}
\title{The quantum geometry of the chemical bond}

\author{\parbox{\dimexpr\linewidth-2\tabcolsep\relax}{\centering\normalsize
Gabrielle Carrel$^{1}$, Nishchhal Verma$^{2}$, Fatmag\"ul Katmer$^{1}$,\\[3pt]
Leslie M. Schoop$^{1,*}$ and Raquel Queiroz$^{2,*}$\\[7pt]
{\small $^{1}$Department of Chemistry, Princeton University,\\
Princeton, NJ 08544, USA\\[3pt]
$^{2}$Department of Physics, Columbia University,\\
New York, NY 10027, USA}\\[7pt]
{\footnotesize $^{*}$Corresponding authors:\\
\href{mailto:lschoop@princeton.edu}{lschoop@princeton.edu};\\
\href{mailto:raquel.queiroz@columbia.edu}{raquel.queiroz@columbia.edu}}
}}

\date{\today}

\maketitle
\begin{abstract}
Electronic properties of materials depend both on band energies and quantum geometry of electronic states. We relate bonding indices to the quantum geometric tensor (QGT), showing how chemical bonding dictates the momentum-space texture of bands, bridging chemical and physical descriptions of solids.
We show the QGT decomposed into the three center bond index, weighted by their oriented area. Resolving these contributions by orbital pair reveals the donor--acceptor character that distinguish ionic, covalent, hypervalent, and metavalent bonding. We identify the bonding pattern for topology in materials: noncollinear three center bonds, which are the source of Berry curvature in solids. Bond fluctuation can be directly linked to electromagnetic response by resolving their energy spectrum, opening a quantitative route towards the design of physical properties of materials from bonding motifs.
\end{abstract}
\section{Introduction}

Chemical bonding and quantum geometry are two ways to understand electronic behavior in molecules and solids. These descriptions often used in different contexts and communities. Here we show that they are intrinsically connected. Every chemical bond fundamentally shapes the quantum geometric tensor, and the resulting physical properties and responses which can be computed with it. We specifically connect the geometric tensor to typical chemical descriptions: bond orders and electron counts, formalized in the Wiberg and Mayer indices\cite{wiberg1968,mayer1983}, crystal orbital bond indices\cite{mullerCrystalOrbitalBond2021}, and the delocalization indices of Bader's theory\cite{fraderaLewisModelBeyond1999}. These bond motifs affect the low-energy behavior and response of materials through their dependence on electronic wavefunctions, from matrix elements of Coulomb interactions and their coupling to external perturbations. 

Quantum geometry \cite{vermaQuantumGeometryRevisiting2025,yuQuantumGeometryQuantum2025, GaoNagaosaRMP2026} describes how electronic wavefunctions in solids depend on their crystal momentum. The quantum geometric tensor (QGT) \cite{provostRiemannianStructureManifolds1980,vermaQuantumGeometryRevisiting2025}, has two components: the quantum metric which measures how much the wavefunction changes between neighboring momenta, and the Berry curvature which measures the phase it picks up when the momentum is moved around a small loop, per unit area of that loop. Quantum geometry is used to quantify the quantum fluctuations of the electronic polarization in the ground state\cite{restaElectronLocalizationInsulating1999,restaInsulatingStateMatter2011,souzaPolarizationLocalizationInsulators2000}, measuring the variance, or mean-square spread, of the electronic dipole moment. Such fluctuations in the absence of thermal motion or an applied field, connect quantum geometry to optical absorption \cite{souzaPolarizationLocalizationInsulators2000}, governing electronic response. 
Its real part being responsible for dielectric susceptibility \cite{komissarovQuantumGeometricOrigin2024} and optical mass \cite{vermaInstantaneous2025, kruchkov2023}; while its imaginary, antisymmetric part governs the intrinsic anomalous Hall effect and encodes band topology \cite{yuQuantumGeometryQuantum2025,nagaosaAnomalousHallEffect2010,thoulessQuantizedHall1982}. Substantial effort has been recently devoted to measuring the full quantum geometric tensor\cite{balutAbbamonteIXS2026,kangMeasurementsQuantumGeometric2025,gianfrateMeasurementQuantumGeometric2020,tanQuantumMetric2019, verma2025step}. 

In this work, we formally relate chemical bonding to quantum geometry, connecting real-space bonding descriptions with the geometric structure of electronic bands. The connection between bonding and polarizability has long guided chemical intuition. The dielectric theories of Phillips and Van Vechten related covalent and ionic bonding to electronic polarizability and electronegativity\cite{phillipsDielectric1968,vanVechtenQuantumDielectric1969}. Ionic NaCl has a small electronic contribution to the dielectric constant, while covalent diamond responds more strongly\cite{komissarovQuantumGeometricOrigin2024}.  Crystalline phase-change materials with extended electron sharing exhibit enhanced electronic contributions to the dielectric constant and anomalous Born effective charges\cite{shportkoResonantBondingCrystalline2008,wuttigIncipientMetalsFunctional2018a}. More recently, the large response of phase-change materials motivated the metavalent bonding framework\cite{wuttigIncipientMetalsFunctional2018a,wuttigPhaseChangeMaterials2009}. Hypervalent bonding has also been extended from molecules to solid-state networks\cite{tremelSquareNetsMaingroup1987,a.papoianHypervalentBondingOne2000a} and has guided work on square-net topological semimetals\cite{klemenzTopologicalSemimetalsSquareNet2019,klemenzRoleDelocalizedChemical2020a}. The relation between hypervalent and metavalent bonding remains debated\cite{leeElliottHyperbonding2020,manjonherreraElectrondeficientMulticenterBonding2024}. For molecules, Kim and Head-Gordon related derivatives of bond indices along a reaction coordinate to bond polarizability\cite{kimEquivalencePolarizabilityBond2024a}. Extending this connection to solids requires relating fluctuations across individual bonds to the macroscopic electronic response.
We show here that the quantum metric can be written as a sum of bond-index contributions weighted by bond lengths squared, linking it to the spatial extent of the electronic distribution. Berry curvature arises from oriented three-center bond indices weighted by the area enclosed by the three orbitals, connecting it to bond loops and multicenter aromaticity indices\cite{giambiagiMulticenterBondIndices2000,bultinckMulticenterBond2005}. Two-center bonds and linear multicenter bonds, such as triiodide\cite{pimentelTrihalide1951}, contribute to the metric but not the curvature. A Chern number therefore needs bond loops that enclose area, something that can be read from a Lewis structure before any calculation.

We derive the decomposition of the quantum geometric tensor into bond contributions, comparing ionic NaCl, covalent diamond, hypervalent \CsI, and metavalent GeTe. We then resolve the metrics velocity matrix elements to examine how transition-current amplitudes are distributed across bonds, showing what can distinguish ionic and covalent character. Finally, we use the Haldane model\cite{haldaneModelQuantumHall1988} to examine how local bond loops contribute to Berry curvature and combine to produce a nonzero Chern number. This work provides a quantitative route from chemical bonding to electronic response and identifies the bond loops that can support band topology, linking real space structure to topology. 

\section{Bonding as quantum geometry}

\begin{figure}[!htbp]
    \centering
    \includegraphics[width=1\linewidth]{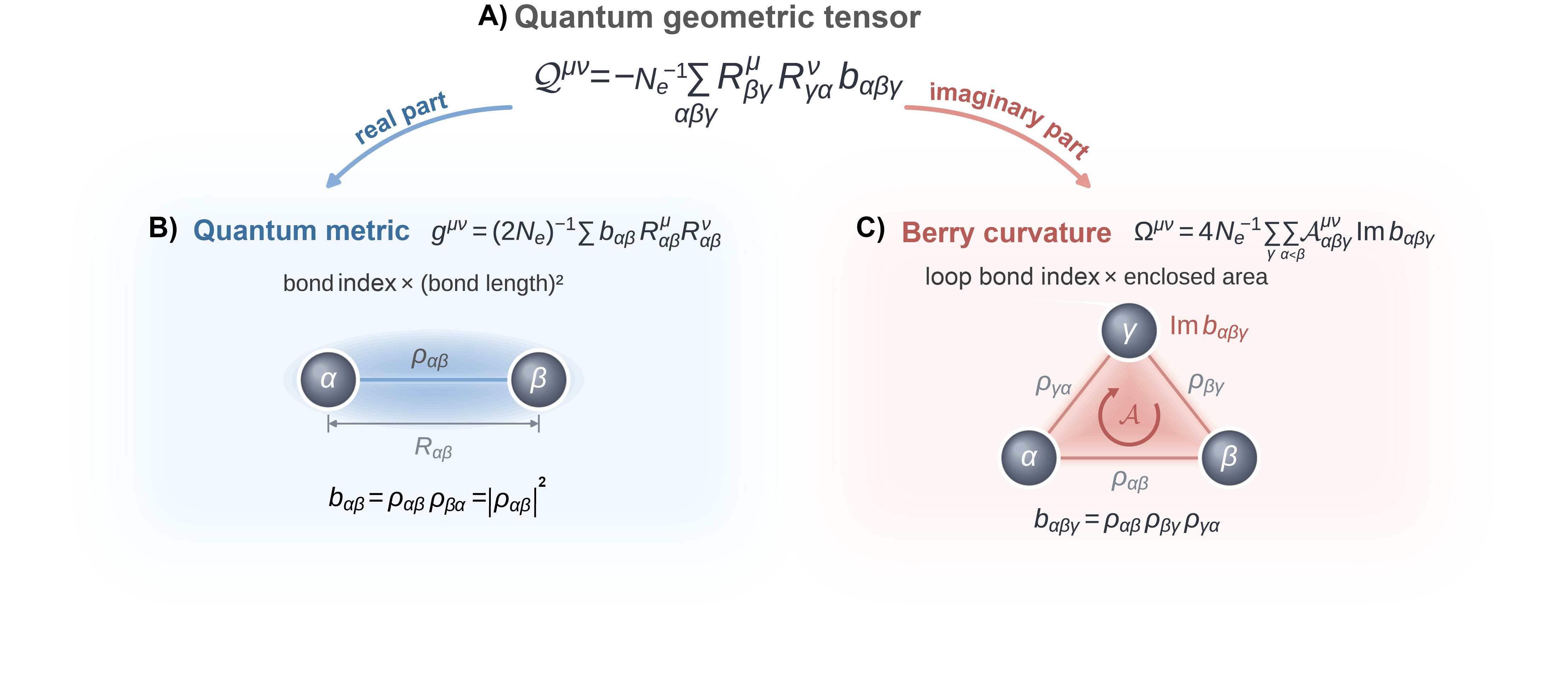}
\caption{\textbf{Chemical bonds build the quantum geometric tensor.}
(\textbf{A}) The QGT built from bonds. Every orbital is placed at the position of its atom, and bond vectors connect the atoms. The QGT is a sum over closed paths through three orbitals, $\alpha\to\beta\to\gamma\to\alpha$. The product of the density-matrix elements along each path gives the three-center bond index $b_{\alpha\beta\gamma}$, weighted by products of bond-vector components.
(\textbf{B}) The tensor's real part, the quantum metric, reduces to a sum over two-center bond indices, $b_{\alpha\beta}=|\rho_{\alpha\beta}|^2$. The metric per electron weights each bond index by bond-vector components, with the normalization of Eq.~\ref{eq:QM_Wiberg_index}. Its trace weights each index by squared bond length, increasing with sharing at fixed length or with length at fixed sharing.
(\textbf{C}) Berry curvature is built from the imaginary part of the three-center bond index, $b_{\alpha\beta\gamma}=\rho_{\alpha\beta}\rho_{\beta\gamma}\rho_{\gamma\alpha}$, weighted by the oriented area enclosed by the three orbitals $\mathcal A^{\mu\nu}_{\alpha\beta\gamma}=\tfrac12(R^\mu_{\beta\gamma}R^\nu_{\gamma\alpha}-R^\nu_{\beta\gamma}R^\mu_{\gamma\alpha})$. This product retains the electronic phase around the loop.
Here $\rho_{\alpha\beta}$ is a matrix element of the occupied-state density matrix and $\mathbf R_{\alpha\beta}=\mathbf r_\alpha-\mathbf r_\beta$ joins the two orbital centers.}
    \label{fig:bond_ladder}
\end{figure}

To describe chemical bonding, we use the one-particle reduced density matrix in an orthonormal, atom-centered basis $|\phi_\alpha\rangle$, with $\alpha$ being a composite label for the atom, orbital, and spin. For an independent-electron ground state, let $|\psi_n\rangle$ and $|\psi_m\rangle$ denote occupied and unoccupied electronic states, respectively. The occupied-state projector is $P=\sum_{n\in\mathrm{occ.}}|\psi_n\rangle\langle\psi_n|$, and its matrix elements $\rho_{\alpha\beta}=\langle\phi_\alpha|P|\phi_\beta\rangle$ describe orbital populations and electron sharing. We compute these elements for the crystalline examples using \textit{ab initio} methods (see materials and methods, section~\ref{app:materials_benchmarks}). The diagonal element $\rho_{\alpha\alpha}$ gives the population of orbital $\alpha$. The off-diagonal coherence element $\rho_{\alpha\beta}$, the Coulson bond order\cite{coulsonFractionalBondOrder1939}, characterizes the phase coherent electron sharing between orbitals $\alpha$ and $\beta$. The squared magnitude, $b_{\alpha\beta}=|\rho_{\alpha\beta}|^2$, gives the orbital contribution to the Wiberg bond index\cite{wiberg1968} (see supplementary text, section~\ref{app:normalization}). In the conventional chemical normalization, this bond order estimates the number of shared electron pairs: close to one for a single covalent bond and close to zero for nonbonding closed-shell contacts. Multiplying the bond orders around a loop of three orbitals,
$b_{\alpha\beta\gamma}=\rho_{\alpha\beta}\rho_{\beta\gamma}\rho_{\gamma\alpha}$
(Fig.~\ref{fig:bond_ladder}\textbf{C}), gives the three-center bond index\cite{sannigrahiThreeCentreBond1988,giambiagiDefinitionMulticenterBond1990}, positive for a three-center two-electron (3c 2e) bond and negative for a three-center four-electron (3c 4e) bond\cite{sannigrahiThreeCentreBond1988}. Conventional multicenter indices\cite{giambiagiDefinitionMulticenterBond1990,giambiagiMulticenterBondIndices2000,bultinckMulticenterBond2005} average over opposite loop orientations. The oriented product instead retains a phase independent of orbital phase conventions, linking bonding to Berry curvature.

The QGT is built from the transition dipole moments between occupied and empty (virtual) orbitals\cite{souzaPolarizationLocalizationInsulators2000,vermaInstantaneous2025}. These transitions are captured by the operator $\mathcal D^\mu=i[\hat r^\mu,P]$ whose matrix elements between filled and empty states are the transition dipole moments per unit charge. Taking the expectation value over occupied electrons, $\langle\cdot\rangle=N_e^{-1}\operatorname{Tr}(P\,\cdot)$ with $N_e$ is the number of electrons, gives the QGT as $\mathcal Q^{\mu\nu}=\langle\mathcal D^\mu\mathcal D^\nu\rangle$ where $\mu$ and $\nu$ are spatial indices. Written out, $\mathcal Q^{\mu\nu}=N_e^{-1}\sum_{n\in\mathrm{occ.}}\sum_{m\in\mathrm{unocc.}}r_{nm}^\mu r_{mn}^\nu$, where $n\in\mathrm{occ.}$ and $m\in\mathrm{unocc.}$ label occupied and unoccupied states, respectively, and $r_{nm}^\mu=\langle\psi_n|\hat r^\mu|\psi_m\rangle$ is the position matrix element between them. This expression applies to molecules and infinite periodic solids (see supplementary text, section~\ref{app:normalization}).

We work in the orbital-center approximation \cite{grafVoglElectromagnetic1995} and an independent-electron description, allowing the quantum geometric tensor to be expressed through bond indices and orbital separations. In this approximation, we place every orbital $\phi_\alpha$ at the position of its atom $r_\alpha$, thus the position operator is $\hat r^{\mu}\simeq\sum_\alpha r^{\mu}_\alpha\Pi_\alpha$, with $\Pi_\alpha=|\phi_\alpha\rangle\langle\phi_\alpha|$ the projector onto orbital $\alpha$ and the sum running over all orbital and spin basis states, including all unit cells of a periodic solid; translated copies have different indices. The dipole matrix elements are then $\langle\phi_\alpha|\mathcal D^\mu|\phi_\beta\rangle=iR^\mu_{\alpha\beta}\rho_{\alpha\beta}$, with $R^\mu_{\alpha\beta}=r^\mu_\alpha-r^\mu_\beta$ the bond separation along direction $\mu$. These elements arise from electron sharing between distinct atoms, the aspect of polarization described by chemical bonding. Substituting into the QGT, we find 
$\mathcal Q^{\mu\nu}=-N_e^{-1}\sum_{\alpha\beta\gamma}R^\mu_{\beta\gamma}R^\nu_{\gamma\alpha}\,b_{\alpha\beta\gamma}$. The sums run over all orbitals. The tensor is thus expressed as a sum of bonding correlations weighted by the vectors connecting orbital centers. Its real part is the localization tensor\cite{restaElectronLocalizationInsulating1999,restaInsulatingStateMatter2011,souzaPolarizationLocalizationInsulators2000,marzariMaximallyLocalizedGeneralized1997}, $g^{\mu\nu}=\operatorname{Re}\mathcal Q^{\mu\nu}$,

\begin{equation}
    g^{\mu\nu} = \frac{1}{2N_e}\sum_{\alpha\beta} b_{\alpha\beta}\, R^\mu_{\alpha\beta} R^\nu_{\alpha\beta},
    \label{eq:QM_Wiberg_index}
\end{equation}
which shows that the metric trace, ${\rm tr}g$, increases with electron sharing at fixed bond length, or with bond length at fixed bond index. The imaginary part, which is antisymmetric in $\mu$ and $\nu$, determines the Berry curvature per electron averaged over the Brillouin zone,
\begin{equation}
    \Omega^{\mu\nu} = -2\,\mathrm{Im}\,\mathcal Q^{\mu\nu} = \frac{4}{N_e}\sum_\gamma\sum_{\alpha<\beta} \mathcal A^{\mu\nu}_{\alpha\beta\gamma}\, \mathrm{Im}\, b_{\alpha\beta\gamma},
    \label{eq:Omega_three_center_main}
\end{equation}
with $\mathcal A^{\mu\nu}_{\alpha\beta\gamma}=\tfrac12(R^\mu_{\beta\gamma}R^\nu_{\gamma\alpha}-R^\nu_{\beta\gamma}R^\mu_{\gamma\alpha})$ the oriented area of the triangle formed by the three atomic sites in the $\mu\nu$ plane. For a two-dimensional periodic insulator this sum determines the quantized Chern number\cite{thoulessQuantizedHall1982} $C=2\pi n_e\Omega^{xy}$ with electron density $n_e$. Therefore, a topological material with strong topological invariants \cite{bradlynTQC2017} requires non-collinear bonds forming loops that enclose a finite area and carry phases whose oriented contributions do not cancel. Such invariants can only be nonzero in a periodic solid. For a finite lattice or a molecule, the sum over all loops is exactly zero because position is a bounded operator (see supplementary text, section~\ref{app:molecular}). 
This is the structural content of band topology in chemical terms. A molecule, aromatic or not, cannot have a Chern number, because every bond loop no matter how strong is eventually compensated by the others. In contrast, in a crystal, loops can add constructively to a topologically invariant. When the boundaries are opened, such as in a finite flake with a boundary, delocalized bonds across the boundary must compensate the bulk loops, forming the topologically protected boundary modes.

\section[Bond fluctuations]{Bond fluctuations}

\label{sec:time_optical_response}

\begin{figure}[!htbp]
\centering
\includegraphics[width=\linewidth]{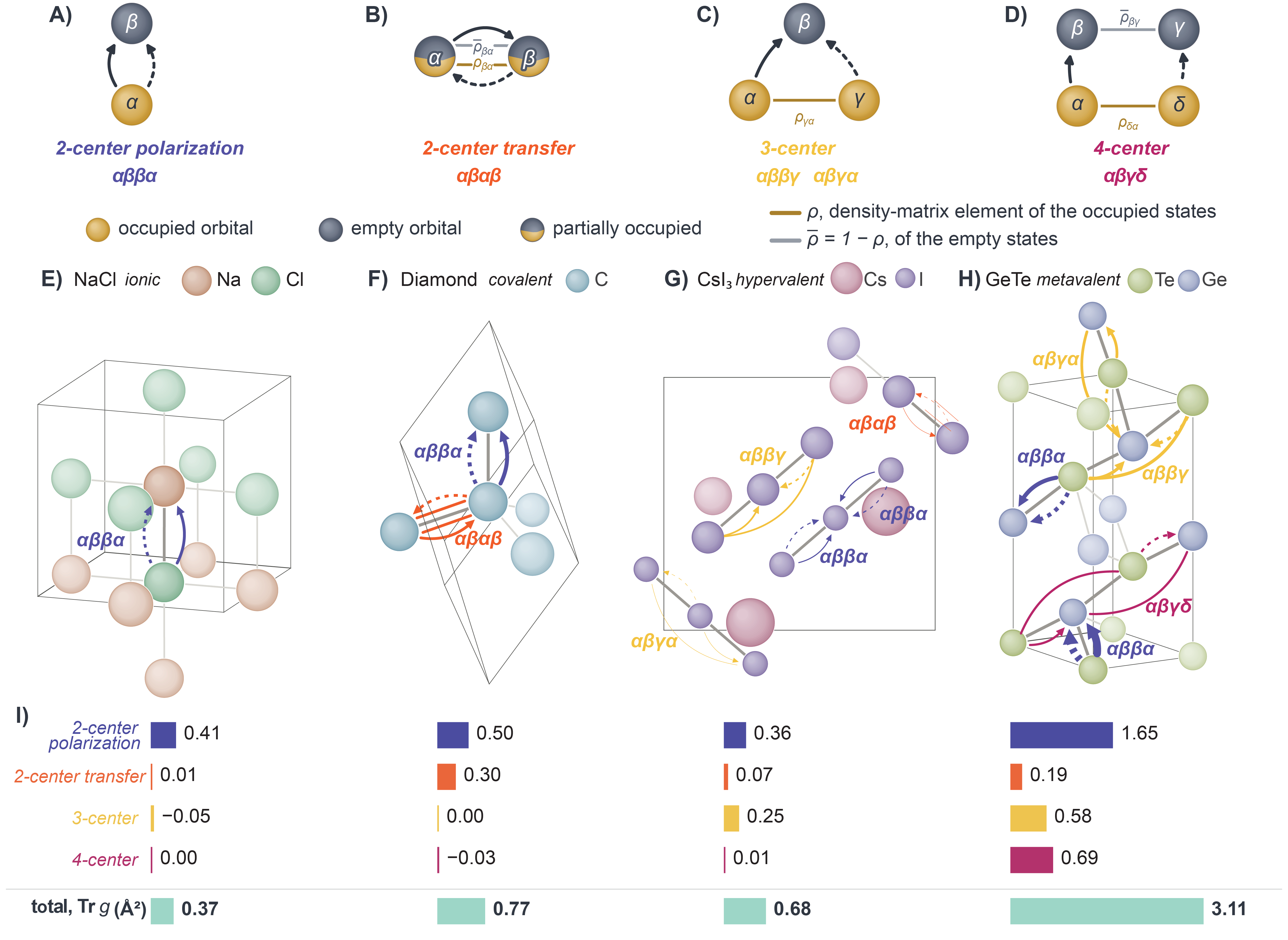}
\caption{\textbf{Bond contributions to electronic fluctuations.}
(\textbf{A--D}) Representative fluctuation patterns for two-center polarization ($\alpha\beta\beta\alpha$), two-center transfer ($\alpha\beta\alpha\beta$), three-center ($\alpha\beta\beta\gamma$), and four-center ($\alpha\beta\gamma\delta$) contributions. For the virtual transition from occupied $\psi_n$ to empty $\psi_m$, the solid arrow $\delta\to\gamma$ represents $\langle\psi_m|v^\nu_{\gamma\delta}|\psi_n\rangle$. The dashed arrow $\alpha\to\beta$ represents $\langle\psi_m|v^\mu_{\beta\alpha}|\psi_n\rangle$, which enters Eq.~\ref{eq:bond_resolved_velocity_tensor_main} complex conjugated as $\langle\psi_n|v^\mu_{\alpha\beta}|\psi_m\rangle$. Connecting lines represent elements of the occupied-state density matrix $\rho$ and its empty-state complement $\bar\rho=1-\rho$, indicating coherent sharing of the filled and empty states across the orbitals, respectively.
(\textbf{E--H}) Fluctuation diagrams of ionic NaCl, covalent diamond, hypervalent CsI, and metavalent GeTe. Line widths scale linearly with contribution to the total metric, colored by the pattern of fluctuation. Contribution to the metric is calculated as the sum of a given fluctuation over all its symmetry-equivalent copies as shown in Table \ref{tab:fig2-fluctuations}.
(\textbf{I}) The contributions regrouped by fluctuation pattern, with the total metric trace in \AA$^2$ below. Different types of bonding show distinct fluctuation patterns, from solely incoherent polarization in NaCl, to long, coherent, multicenter fluctuations in GeTe.}

\label{fig:bonding_combined}
\end{figure}

The bond-index decomposition relates electronic fluctuations to electron sharing and the distances between orbitals. We now ask whether the tensor also contains information about the bonding motifs that chemists distinguish: ionic, covalent, hypervalent, and metavalent. To examine these fluctuations, we express the QGT through velocity matrix elements connecting occupied and empty states, which reveal the directionality of the bond and the donor and acceptor orbitals.

A virtual transition connects an occupied $|\psi_n\rangle$ to an unoccupied  $|\psi_m\rangle$ state, separated in energy by $\epsilon_m-\epsilon_n$. Ground-state quantum fluctuations in general involve multiple orbitals because the electron states $|\psi_n\rangle$ and $|\psi_m\rangle$ are often coherent superpositions of multiple orbitals. The current matrix element is $-e$ times the velocity matrix element, which we resolve between pairs of atomic orbitals, $\langle\phi_\alpha|v^\mu|\phi_\beta\rangle=-(i/\hbar)H_{\alpha\beta}R^\mu_{\alpha\beta}$: the resonance integral between the orbitals multiplied by their separation along direction $\mu$. This connects the decomposition to the hopping amplitudes familiar from H\"uckel theory\cite{huckelBenzolproblem1931,grafVoglElectromagnetic1995}. Writing $\left\langle\psi_n\right| \mathcal{D}^\mu\left|\psi_m\right\rangle={\hbar\left\langle\psi_n\right| v^\mu\left|\psi_m\right\rangle}/{(\epsilon_m-\epsilon_n)}$  (see supplementary text, section~\ref{app:decomposition_equivalence}), we can then decompose the QGT into bond current fluctuations $\mathcal Q^{\mu\nu}=\sum_{\alpha\beta\gamma\delta}Q^{\mu\nu}_{\alpha\beta\gamma\delta}$, with
\begin{equation}
 Q^{\mu\nu}_{\alpha\beta\gamma\delta}
 =\frac{\hbar^2}{N_e}
 \sum_{\substack{n\in\mathrm{occ.}\\m\in\mathrm{unocc.}}}
 \frac{
 \langle\psi_n|v^\mu_{\alpha\beta}|\psi_m\rangle
 \langle\psi_m|v^\nu_{\gamma\delta}|\psi_n\rangle
 }{(\epsilon_m-\epsilon_n)^2}.
 \label{eq:bond_resolved_velocity_tensor_main}
\end{equation}

The four labels identify the two orbital pairs contributing to one excitation. Their relative phases determine whether the contributions reinforce or cancel. Figure~\ref{fig:bonding_combined}\textbf{A--D} illustrates four types of contributions. We use the relative weights and spatial organization of these fluctuation types as proxies for bonding types. (i) \emph{Two-center polarization}, $\alpha\beta\beta\alpha$, pairs the directed transfer, $\alpha\to\beta$, with itself. The occupied state has weight on the donor orbital and the empty state on the acceptor orbital. This is the simplest polarization pattern, common to all four bonding types. (ii) \emph{Two-center transfer}, $\alpha\beta\alpha\beta$, describes interference between transfers in opposite directions on the same orbital pair. Both orbitals have weight in the occupied and empty states, as in a transition from a bonding to an antibonding orbital. (iii) \emph{Three-center} terms probe electron sharing across three distinct atomic centers, as in the hypervalent iodine trimer\cite{pimentelTrihalide1951}. In the illustrated pattern $\alpha\beta\beta\gamma$, donors $\alpha$ and $\gamma$ share acceptor $\beta$. Their relative phase reflects delocalization of the occupied state across the donors. All three-center arrangements, including those with one donor and two acceptors, form one class (see supplementary text, section~\ref{app:class_physics}). (iv) \emph{Four-center} terms, $\alpha\beta\gamma\delta$, couple transfers $\alpha\to\beta$ and $\delta\to\gamma$ across four distinct centers. They probe phase relationships in both occupied and empty states and electron sharing through the metavalent GeTe network. Together these patterns describe polarization and coherent electron sharing across several centers and indicate bonding character without assigning any pattern to a single bond type.

\begin{figure}[tbp]
\centering
\includegraphics[width=\linewidth]{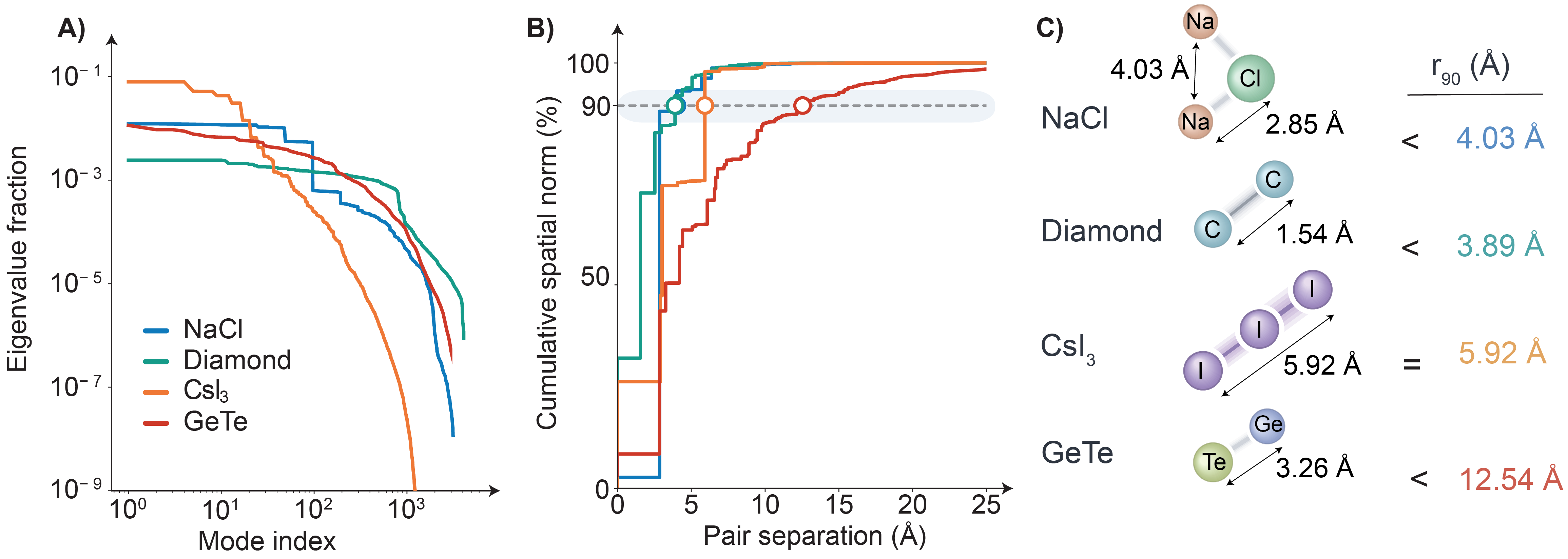}

\caption{\textbf{Fluctuation spectra and spatial range.}
(\textbf{A}) Eigenvalues of the full bond-bond fluctuation matrix, normalized by their sum. The full response has orthogonal eigenmodes, but a fluctuation motif can combine several of them.
Their normalized transition kernels give the probability associated with each ordered pair of atomic sites and relative lattice displacement.
(\textbf{B}) Cumulative weight of each selected fluctuation as a function of the distance between the two atomic sites. The circles mark $r_{90}$, the distance enclosing $90\%$ of its normalized weight; the curves include the full distribution.
(\textbf{C}) Comparison of $r_{90}$ with representative atomic separations and the iodine-trimer length. The displayed ranges are $4.03$, $3.89$, $5.92$ and $12.54$~\AA\ for NaCl, diamond, \CsI\ and GeTe, respectively.  The global spectrum in (\textbf{A}) is distinct from the selected fluctuations in (\textbf{B}), reconstructed from the spectral group contributing most to the metric within each chosen bonding unit, including correlations with the surrounding crystal (see materials and methods, section~\ref{app:physical_image_channel_modes_v15}).}
\label{fig:linked_channel_modes_v15}
\end{figure}

We now display this information for ionic NaCl, covalent diamond, hypervalent \CsI, and metavalent GeTe in two complementary ways: through atom-pair bonding patterns and their contributions (Fig.~\ref{fig:bonding_combined}\textbf{E--H}), and through fluctuation spectra and spatial profiles (Fig.~\ref{fig:linked_channel_modes_v15}). For these time-reversal-symmetric materials, the Brillouin-zone-averaged Berry curvature vanishes, so we compare contributions to the metric trace (Fig.~\ref{fig:bonding_combined}\textbf{I}).
In ionic NaCl, virtual transitions predominantly transfer electron density from chloride toward neighboring sodium ions. These are fluctuations of localized orbitals about the ionic ground state, conventionally described as Na$^+$ and Cl$^-$. This donor--acceptor asymmetry is consistent with ionic bonding. In covalent diamond, the fluctuations are also concentrated on nearest-neighbor bonds, but the filled states (bonding orbitals) are coherently delocalized over the neighboring carbon atoms, resulting in phase coherent transfer fluctuations from bonding to anti-bonding states, consistent with covalent bonding. In hypervalent \CsI, the fluctuations are concentrated within the I$_3^-$ trimer , with the terminal iodines acting predominantly as donors and the central iodine as acceptor. The three-center term pairs the two terminal-to-central transfers through their common central iodine, connecting the fluctuation to the three-center bonding motif. The retained model contains iodine orbitals only, so contributions involving cesium are not evaluated. In GeTe, the fluctuations involve both short and long contacts\cite{wuttigIncipientMetalsFunctional2018a,leeElliottHyperbonding2020}, revealing coherent electron sharing distributed across the surrounding bonding network. The four fluctuation types encode the distinction between ionic, covalent, and hypervalent and metavalent bonding, linking qualitative bonding types and electronic fluctuations.  

We next examine the spectrum and spatial range of fluctuations (Fig.~\ref{fig:linked_channel_modes_v15}). We view the four-index object in Eq.~\ref{eq:bond_resolved_velocity_tensor_main} as a bond-bond fluctuation matrix and perform an eigendecomposition to get the fluctuation spectrum and normal modes. 
For each material we follow the fluctuation of that contributes most to the metric, a Na--Cl pair, a carbon with its neighbors, an I$_3^-$ unit, or a germanium with its neighbors, into the crystal, and record $r_{90}$, the distance holding $90\%$ of its weight (see materials and methods, section~\ref{app:physical_image_channel_modes_v15}). The fluctuation in hypervalent \CsI\ extends across the iodine trimer, farther than those in NaCl and diamond. The selected fluctuation in metavalent GeTe is the most delocalized of the four, spanning several bond lengths. The same three-center fluctuation appears in \CsI\ and in GeTe, confined to the I$_3^-$ unit in one and running through the whole crystal in the other.  This comparison does not resolve the broader relationship between hypervalent and metavalent bonding\cite{leeElliottHyperbonding2020,manjonherreraElectrondeficientMulticenterBonding2024}. A comparison with the hypervalent and metavalent descriptions is given in the supplementary text (section~\ref{app:hypervalent_metavalent}).

\section{From bond fluctuations to measurable response}

The QGT measures the spatial extent of charge fluctuations without resolving their energy cost. The fluctuation--dissipation theorem \cite{kuboFluctuationdissipationTheorem} connects these equilibrium fluctuations to optical absorption. Thus, sorting these virtual transitions by their energy $E$ connects the bond picture to spectroscopy: light can drive the corresponding excitations resonantly when its photon energy $\hbar\omega$ matches $E$.  
At $E=\hbar\omega$, the absorptive optical conductivity is $\sigma^{\mathrm{abs},\mu\nu}(\omega)=\pi e^2n\omega\,\mathcal Q^{\mu\nu}(\hbar\omega)$, where $n=N_e/V$ is the electron number density and the energy resolved tensor is given by $\mathcal Q^{\mu\nu}(E)=N_e^{-1}\sum_{n\in\mathrm{occ.}}\sum_{m\in\mathrm{unocc.}}r^\mu_{nm}r^\nu_{mn}\,\delta[E-(\epsilon_m-\epsilon_n)]$.  (see supplementary text, section~\ref{app:correlator_derivation}). Its real symmetric part describes ordinary absorption; its imaginary antisymmetric part describes magnetic circular dichroism, the difference in absorption of opposite circular polarizations, which requires broken time reversal in this electric-dipole response. Because the transition amplitudes are sums of bond contributions, the conductivity can also be resolved bond by bond (Eq.~\ref{eq:bond_resolved_velocity_tensor_main}).
Integrating $\mathcal{Q}^{\mu\nu}(E)$ with different energy weights, gives familiar quantities in terms of sum rules \cite{vermaInstantaneous2025, Souza2025}. Weighting the real part by $1/E$ gives the static electronic polarizability of second-order perturbation theory and, with the electron density, the electronic dielectric constant \cite{vermaInstantaneous2025,vermaFrameworkMeasureQuantum2025}. Weighting it by $E$ gives the oscillator-strength sum, or equivalently the inverse optical mass. The polarizability therefore favors fluctuations that are both large and cheap. This provides a physical picture for the large electronic dielectric response of metavalent GeTe: its spatially extended fluctuations are especially effective at polarizing the material when they occur at low excitation energies\cite{shportkoResonantBondingCrystalline2008,wuttigIncipientMetalsFunctional2018a}. 
The imaginary part of the energy-weighted integral gives the self-rotation contribution to the orbital magnetic moment through the dichroic sum rule\cite{souzaDichroic$f$sumRule2008,restaXrayCircularDichroism2020}. In a molecule, the orbital moment is the sum of $\mathbf r \times \mathbf I$ over the bond currents induced by a static magnetic field or by broken time-reversal symmetry, as for the H\"uckel--London currents circulating around molecular rings \cite{london1937,pople1958}. In a crystal, however, the bond currents within a cell determine only the self-rotation part, while the itinerant part requires a boundary \cite{ceresoliOrbitalMagnetization2006, bianco2013}.
The antisymmetric spectrum also connects to static response since Kramers--Kronig relations relate magnetic circular dichroism to the dc Hall conductivity in an insulator \cite{souzaDichroic$f$sumRule2008}. This brings us back to closed loops of bonds, whose geometric contributions we examine next.

\section{Loop chemistry and topology}

\begin{figure*}
\centering
\includegraphics[width=0.86\linewidth]{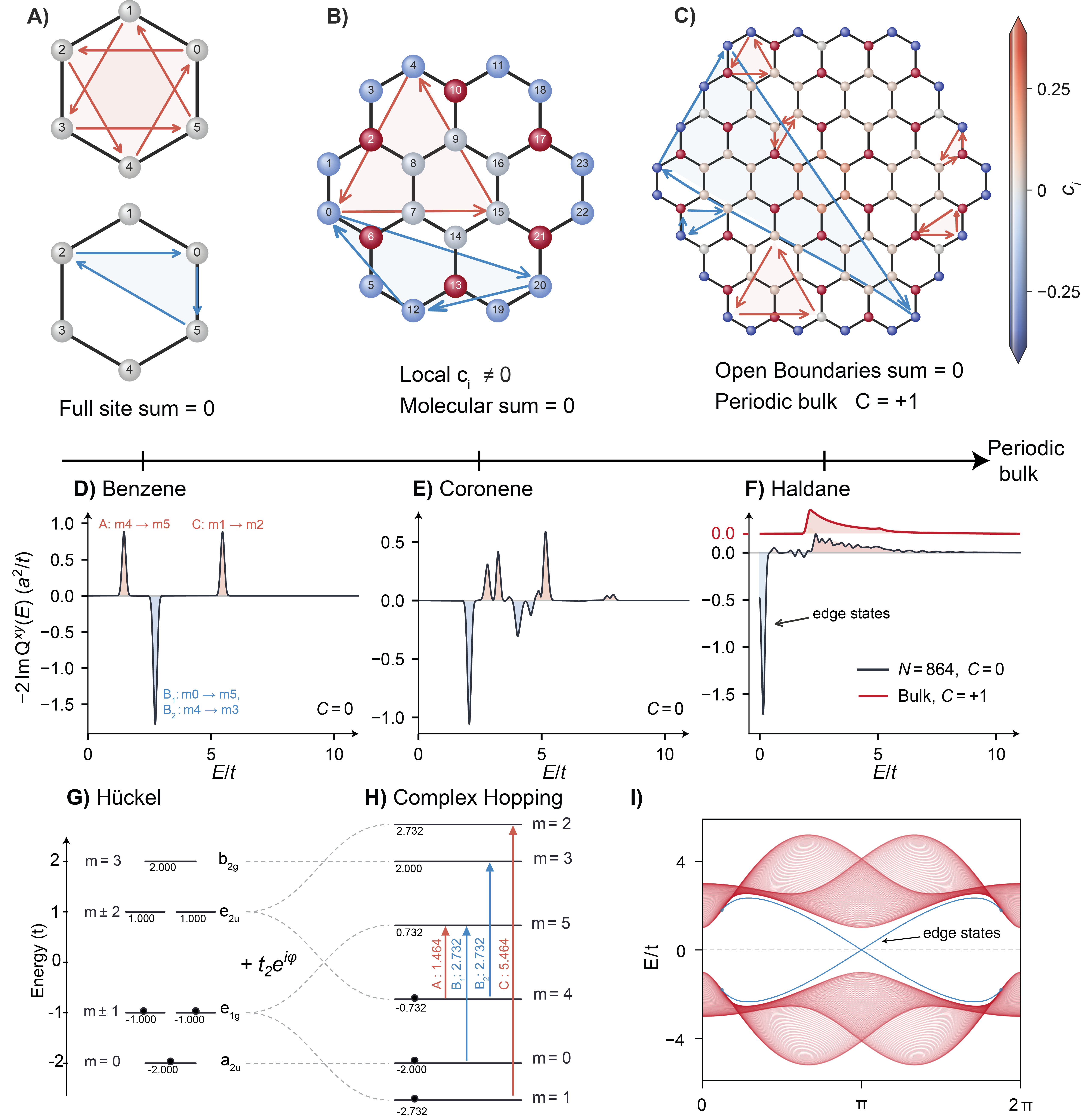}
\caption{\textbf{From bond loops to the Chern number of a crystal.}
(\textbf{A--C}) Signed geometric loop contributions on benzene, coronene and a 96-site flake; arrows denote directed bonds. Normalized site sums obey $\sum_i c_i=C$ for the periodic sample; complete finite-system sums vanish.
(\textbf{D--F}) Curvature spectra $-2\operatorname{Im}\mathcal Q^{xy}(E)$ per electron, in $a^2/t$. Red minus blue area gives the loop sum of Eq.~\ref{eq:Omega_three_center_main}. Multiplying by excitation energy gives the transverse absorptive conductivity, with prefactors in the supplementary text (section~\ref{app:correlator_derivation}). In benzene and coronene, the positive and negative spectral areas cancel. In (\textbf{F}), the periodic bulk (red, offset) has $C=+1$; the 864-site flake (black) has an edge peak just above zero energy cancelling its bulk area. Red and blue denote positive and negative contributions about each curve's baseline.
(\textbf{G,H}) H\"uckel levels split by complex hopping, with opposite circular-polarization transitions.
(\textbf{I}) Zigzag ribbon: staggered flux opens a topological gap with blue edge branches. Broken time reversal produces unequal circular-polarization strengths.
All geometries use the Hamiltonian with $t'=t$, $\phi=\pi/2$, equal sublattice energies preserving bulk inversion; only the geometry and boundary conditions change. Molecular graphs are models with complex hopping, not zero-field molecules. Here $a$ is the nearest-neighbor distance. The level inversion in (\textbf{H}) which leads to resonant states of negative chirality ($B_1$ and $B_2$) is the molecular precursor of the topological edge modes that compensate the net positive bulk Hall response in the Haldane model (\textbf{I}).}
\label{fig:loop_chemistry_summary}
\end{figure*}

Ring currents in aromatic molecules are familiar to chemists through their contribution to proton NMR shifts\cite{paulingDiamagneticAnisotropy1936,popleRingCurrents1956,schleyerNICS1996,gershoniporanneNICS2015}. The field-induced phases in the hopping matrix elements also allow complex three-center bond products. Their imaginary parts, weighted by the enclosed oriented areas, give the loop contributions to the curvature of Eq.~\ref{eq:Omega_three_center_main}. Summing these contributions lets us ask which loops cancel and which survive, and how this differs between a finite molecule and a periodic crystal. For a spinless, time-reversal-symmetric H\"uckel model, the density matrix is real, making every three-center product $b_{\alpha\beta\gamma}$ real. In this case, no loop contributes an imaginary part to the curvature. To study a loop with a phase, we use a toy model, the spinless Haldane model\cite{haldaneModelQuantumHall1988}, on a honeycomb net with one orbital per site (see materials and methods, section~\ref{app:tb_models}). The bond and fluctuation decompositions of the same three honeycomb models, from ring to net, are shown in fig.~\ref{fig:SI_benzene_fluctuation} (supplementary text, section~\ref{sec:honeycomb_fluctuation}). Nearest neighbors are coupled by a real hopping $t$. To mimic time-reversal breaking from intrinsic magnetization or an external field, we introduce complex next-nearest-neighbor hopping\cite{london1937} with magnitude $t'$ and phase $\pm\phi$, with the sign set by the orientation of the two-step path through their common neighbor. This pattern breaks time reversal while keeping the net magnetic flux per unit cell zero. Under this fictitious field, individual triangle contributions are nonzero in all the structures of Fig.~\ref{fig:loop_chemistry_summary}. In benzene, at this large flux, two levels are inverted, creating large compensating loop contributions. The two triangles formed by alternating carbon sites add, but triangles containing adjacent sites cancel their contribution, leaving zero at each site.  In coronene, the local sums can be nonzero even though their total vanishes. Summing the signed, area-weighted triangle contributions associated with a given site gives a quantity proportional to the local Chern marker of Bianco and Resta\cite{biancoMappingTopologicalOrder2011}. We denote the normalized site contributions by $c_i$ (see supplementary text, section~\ref{sec:chern-motifs-v15}). The complete sum vanishes in any finite system because its position operators are bounded and commute; a nonzero bulk marker is compensated by the boundary contributions. In the periodic lattice, the loop sum per cell instead gives the Chern number\cite{biancoMappingTopologicalOrder2011,fukuiChernNumbers2005}.

A vanishing integrated curvature does not imply an absence of circular dichroism. Complex hopping redistributes the strengths of opposite circular-polarization transitions across excitation energies (Fig.~\ref{fig:loop_chemistry_summary}\textbf{D--H})\cite{souzaDichroic$f$sumRule2008,restaXrayCircularDichroism2020}. A level inversion such as one caused in benzene by a large flux, depicted in Fig.~\ref{fig:loop_chemistry_summary}\textbf{G,H}, leads to a large $\mathcal{Q}^{xy}(E)$ response reflected in chiral circulating currents that can be resonantly excited. These compensating modes evolve to the coronene, and finally, in the finite flake, become the chiral edge states that supply the compensating optical response required by the boundary. In {Fig.~\ref{fig:loop_chemistry_summary}\textbf{A -- B}} we can see how such compensating fluctuations appear close to the boundary in the finite molecules. 

At the parameters used here, nearly half of the Chern number ($\approx42\%$) comes from one family of triangles: two nearest-neighbor bonds closed by one second-neighbor contact across the two sublattices (fig.~\ref{fig:motifs-haldane-v15}\textbf{A}). The triangles of alternating atoms, which add in the benzene graph, cancel in the periodic sum at these parameters through translation symmetry and an identity of the density matrix of the infinite lattice. The remaining families and the cancellation identities are given in the supplementary text (section~\ref{sec:chern-motifs-v15}). The loop analysis applies also to real crystals. With spin--orbit coupling, the occupied states can be complex even when time reversal is preserved. Their time-reversed Kramers partners then cancel in the total charge curvature, without requiring a conserved spin. The septuple layers of MnBi$_2$Te$_4$ combine a Bi--Te framework with Mn magnetic order and strong spin--orbit coupling; suitable thin films exhibit the quantum anomalous Hall effect\cite{otrokovUniqueThicknessDependent2019,dengQuantumAnomalousHall2020}. A  simple Lewis structure is therefore the first test of a candidate topological crystal. It shows which loops may carry curvature and the bond phases, defined by magnetism, spin--orbit coupling and local symmetry determine if such material is in fact topological.

\section{Conclusion}

Chemistry and physics describe the electrons in molecules and solids using different languages, as Hoffmann emphasized in ``How chemistry and physics meet in the solid state''\cite{hoffmannHowChemistryPhysics1987}. Here we have shown that the two languages describe the same thing. The quantum metric is a sum of bond indices times the bond length squared, and the Berry curvature is a sum of bond loops multiplied by the area of the loop. Thus, both terms of the quantum geometric tensor, a tensor that describes the dielectric properties, absorption spectrum and Hall responses of materials, are related to drawable objects of Lewis structures. This relationship allows us to distinguish different types of chemical bonding in four exemplary solids, becoming a natural proxy to chemically describe solids. In addition, we recognize design principles for topology in the spatial structure of orbitals. For a nonzero Chern number, noncollinear bond loops are required, in addition to magnetism or spin--orbit coupling who are responsible for level inversion that exchange filled and empty states, thus generating long range coherence. This effect can be seen already in molecules. However, the phase that arises from bond loops must always cancel in molecules, while it may remain in a solid, where it then must be compensated by topologically protected edge modes. These modes arise from coherent electron sharing along the edge, as seen in our formalism. In the Haldane lattice, the largest share of the Chern number comes from one family of triangles, which provides additional hints for the discovery of new topological materials. Time-reversal-invariant, or non-magnetic topological insulators can remain topological even though the Berry-curvature contributions of time-reversed partners cancel in the total\cite{hasanColloquiumTopologicalInsulators2010a}. Next it should be tested whether analogy derived here can be extend to metals, where bond indices decay slowly, and correlated solids, where the reduced density matrix no longer describes the full chemical picture.

\section{Acknowledgments}
R.Q. acknowledges support from the Alfred P. Sloan Foundation (FG-2025-24714) and the National Science Foundation through the CAREER Award (DMR-2340394). L.M.S. acknowledges support from the Air Force Office of Scientific Research through a Presidential Early Career Award for Scientists and Engineers under award number FA9550-25-1-0177. This research was supported by Grant No.~2024178 from the United States-Israel Binational Science Foundation (BSF). We acknowledge Princeton Research Computing, a consortium including the Princeton Institute for Computational Science and Engineering (PICSciE) and Research Computing at Princeton University, for computational resources. This material is based on work supported by the National Science Foundation Graduate Research Fellowship Program under Grant No. DGE-2444107. Data and materials availability: all data for this work are available on Zenodo under DOI 10.5281/zenodo.23243055.

\newpage
\bibliographystyle{apsrev4-1}
\bibliography{refs_v15}

\begin{thebibliography}{67}\makeatletter
\providecommand \@ifxundefined [1]{ \@ifx{#1\undefined}
}\providecommand \@ifnum [1]{ \ifnum #1\expandafter \@firstoftwo
 \else \expandafter \@secondoftwo
 \fi
}\providecommand \@ifx [1]{ \ifx #1\expandafter \@firstoftwo
 \else \expandafter \@secondoftwo
 \fi
}\providecommand \natexlab [1]{#1}\providecommand \enquote  [1]{``#1''}\providecommand \bibnamefont  [1]{#1}\providecommand \bibfnamefont [1]{#1}\providecommand \citenamefont [1]{#1}\providecommand \href@noop [0]{\@secondoftwo}\providecommand \href [0]{\begingroup \@sanitize@url \@href}\providecommand \@href[1]{\@@startlink{#1}\@@href}\providecommand \@@href[1]{\endgroup#1\@@endlink}\providecommand \@sanitize@url [0]{\catcode `\\12\catcode `\$12\catcode `\&12\catcode `\#12\catcode `\^12\catcode `\_12\catcode `\%12\relax}\providecommand \@@startlink[1]{}\providecommand \@@endlink[0]{}\providecommand \url  [0]{\begingroup\@sanitize@url \@url }\providecommand \@url [1]{\endgroup\@href {#1}{\urlprefix }}\providecommand \urlprefix  [0]{URL }\providecommand \Eprint [0]{\href }\providecommand \doibase [0]{http://dx.doi.org/}\providecommand \selectlanguage [0]{\@gobble}\providecommand \bibinfo  [0]{\@secondoftwo}\providecommand \bibfield  [0]{\@secondoftwo}\providecommand \translation [1]{[#1]}\providecommand \BibitemOpen [0]{}\providecommand \bibitemStop [0]{}\providecommand \bibitemNoStop [0]{.\EOS\space}\providecommand \EOS [0]{\spacefactor3000\relax}\providecommand \BibitemShut  [1]{\csname bibitem#1\endcsname}\let\auto@bib@innerbib\@empty
\bibitem [{\citenamefont {Wiberg}(1968)}]{wiberg1968}  \BibitemOpen
  \bibfield  {author} {\bibinfo {author} {\bibfnamefont {K.~B.}\ \bibnamefont {Wiberg}},\ }\href {\doibase 10.1016/0040-4020(68)88057-3} {\bibfield  {journal} {\bibinfo  {journal} {Tetrahedron}\ }\textbf {\bibinfo {volume} {24}},\ \bibinfo {pages} {1083} (\bibinfo {year} {1968})}\BibitemShut {NoStop}\bibitem [{\citenamefont {Mayer}(1983)}]{mayer1983}  \BibitemOpen
  \bibfield  {author} {\bibinfo {author} {\bibfnamefont {I.}~\bibnamefont {Mayer}},\ }\href {\doibase 10.1016/0009-2614(83)80005-0} {\bibfield  {journal} {\bibinfo  {journal} {Chemical Physics Letters}\ }\textbf {\bibinfo {volume} {97}},\ \bibinfo {pages} {270} (\bibinfo {year} {1983})}\BibitemShut {NoStop}\bibitem [{\citenamefont {M{\"u}ller}\ \emph {et~al.}(2021)\citenamefont {M{\"u}ller}, \citenamefont {Ertural}, \citenamefont {Hempelmann},\ and\ \citenamefont {Dronskowski}}]{mullerCrystalOrbitalBond2021}  \BibitemOpen
  \bibfield  {author} {\bibinfo {author} {\bibfnamefont {P.~C.}\ \bibnamefont {M{\"u}ller}}, \bibinfo {author} {\bibfnamefont {C.}~\bibnamefont {Ertural}}, \bibinfo {author} {\bibfnamefont {J.}~\bibnamefont {Hempelmann}}, \ and\ \bibinfo {author} {\bibfnamefont {R.}~\bibnamefont {Dronskowski}},\ }\href {\doibase 10.1021/acs.jpcc.1c00718} {\bibfield  {journal} {\bibinfo  {journal} {The Journal of Physical Chemistry C}\ }\textbf {\bibinfo {volume} {125}},\ \bibinfo {pages} {7959} (\bibinfo {year} {2021})}\BibitemShut {NoStop}\bibitem [{\citenamefont {Fradera}\ \emph {et~al.}(1999)\citenamefont {Fradera}, \citenamefont {Austen},\ and\ \citenamefont {Bader}}]{fraderaLewisModelBeyond1999}  \BibitemOpen
  \bibfield  {author} {\bibinfo {author} {\bibfnamefont {X.}~\bibnamefont {Fradera}}, \bibinfo {author} {\bibfnamefont {M.~A.}\ \bibnamefont {Austen}}, \ and\ \bibinfo {author} {\bibfnamefont {R.~F.~W.}\ \bibnamefont {Bader}},\ }\href {\doibase 10.1021/jp983362q} {\bibfield  {journal} {\bibinfo  {journal} {The Journal of Physical Chemistry A}\ }\textbf {\bibinfo {volume} {103}},\ \bibinfo {pages} {304} (\bibinfo {year} {1999})}\BibitemShut {NoStop}\bibitem [{\citenamefont {Verma}\ \emph {et~al.}(2026)\citenamefont {Verma}, \citenamefont {Moll}, \citenamefont {Holder},\ and\ \citenamefont {Queiroz}}]{vermaQuantumGeometryRevisiting2025}  \BibitemOpen
  \bibfield  {author} {\bibinfo {author} {\bibfnamefont {N.}~\bibnamefont {Verma}}, \bibinfo {author} {\bibfnamefont {P.~J.~W.}\ \bibnamefont {Moll}}, \bibinfo {author} {\bibfnamefont {T.}~\bibnamefont {Holder}}, \ and\ \bibinfo {author} {\bibfnamefont {R.}~\bibnamefont {Queiroz}},\ }\href {\doibase 10.1038/s42254-026-00923-y} {\bibfield  {journal} {\bibinfo  {journal} {Nature Reviews Physics}\ }\textbf {\bibinfo {volume} {8}},\ \bibinfo {pages} {226} (\bibinfo {year} {2026})},\ \Eprint {http://arxiv.org/abs/2504.07173} {arXiv:2504.07173} \BibitemShut {NoStop}\bibitem [{\citenamefont {Yu}\ \emph {et~al.}(2025)\citenamefont {Yu}, \citenamefont {Bernevig}, \citenamefont {Queiroz}, \citenamefont {Rossi}, \citenamefont {T{\"o}rm{\"a}},\ and\ \citenamefont {Yang}}]{yuQuantumGeometryQuantum2025}  \BibitemOpen
  \bibfield  {author} {\bibinfo {author} {\bibfnamefont {J.}~\bibnamefont {Yu}}, \bibinfo {author} {\bibfnamefont {B.~A.}\ \bibnamefont {Bernevig}}, \bibinfo {author} {\bibfnamefont {R.}~\bibnamefont {Queiroz}}, \bibinfo {author} {\bibfnamefont {E.}~\bibnamefont {Rossi}}, \bibinfo {author} {\bibfnamefont {P.}~\bibnamefont {T{\"o}rm{\"a}}}, \ and\ \bibinfo {author} {\bibfnamefont {B.-J.}\ \bibnamefont {Yang}},\ }\href {\doibase 10.1038/s41535-025-00801-3} {\bibfield  {journal} {\bibinfo  {journal} {npj Quantum Materials}\ }\textbf {\bibinfo {volume} {10}},\ \bibinfo {pages} {101} (\bibinfo {year} {2025})}\BibitemShut {NoStop}\bibitem [{\citenamefont {Gao}\ \emph {et~al.}(2026)\citenamefont {Gao}, \citenamefont {Nagaosa}, \citenamefont {Ni},\ and\ \citenamefont {Xu}}]{GaoNagaosaRMP2026}  \BibitemOpen
  \bibfield  {author} {\bibinfo {author} {\bibfnamefont {A.}~\bibnamefont {Gao}}, \bibinfo {author} {\bibfnamefont {N.}~\bibnamefont {Nagaosa}}, \bibinfo {author} {\bibfnamefont {N.}~\bibnamefont {Ni}}, \ and\ \bibinfo {author} {\bibfnamefont {S.-Y.}\ \bibnamefont {Xu}},\ }\href {\doibase 10.1103/f78t-ky69} {\bibfield  {journal} {\bibinfo  {journal} {Rev. Mod. Phys.}\ }\textbf {\bibinfo {volume} {98}},\ \bibinfo {pages} {035005} (\bibinfo {year} {2026})}\BibitemShut {NoStop}\bibitem [{\citenamefont {Provost}\ and\ \citenamefont {Vallee}(1980)}]{provostRiemannianStructureManifolds1980}  \BibitemOpen
  \bibfield  {author} {\bibinfo {author} {\bibfnamefont {J.~P.}\ \bibnamefont {Provost}}\ and\ \bibinfo {author} {\bibfnamefont {G.}~\bibnamefont {Vallee}},\ }\href {\doibase 10.1007/BF02193559} {\bibfield  {journal} {\bibinfo  {journal} {Communications in Mathematical Physics}\ }\textbf {\bibinfo {volume} {76}},\ \bibinfo {pages} {289} (\bibinfo {year} {1980})}\BibitemShut {NoStop}\bibitem [{\citenamefont {Resta}\ and\ \citenamefont {Sorella}(1999)}]{restaElectronLocalizationInsulating1999}  \BibitemOpen
  \bibfield  {author} {\bibinfo {author} {\bibfnamefont {R.}~\bibnamefont {Resta}}\ and\ \bibinfo {author} {\bibfnamefont {S.}~\bibnamefont {Sorella}},\ }\href {\doibase 10.1103/PhysRevLett.82.370} {\bibfield  {journal} {\bibinfo  {journal} {Physical Review Letters}\ }\textbf {\bibinfo {volume} {82}},\ \bibinfo {pages} {370} (\bibinfo {year} {1999})}\BibitemShut {NoStop}\bibitem [{\citenamefont {Resta}(2011)}]{restaInsulatingStateMatter2011}  \BibitemOpen
  \bibfield  {author} {\bibinfo {author} {\bibfnamefont {R.}~\bibnamefont {Resta}},\ }\href {\doibase 10.1140/epjb/e2010-10874-4} {\bibfield  {journal} {\bibinfo  {journal} {The European Physical Journal B}\ }\textbf {\bibinfo {volume} {79}},\ \bibinfo {pages} {121} (\bibinfo {year} {2011})}\BibitemShut {NoStop}\bibitem [{\citenamefont {Souza}\ \emph {et~al.}(2000)\citenamefont {Souza}, \citenamefont {Wilkens},\ and\ \citenamefont {Martin}}]{souzaPolarizationLocalizationInsulators2000}  \BibitemOpen
  \bibfield  {author} {\bibinfo {author} {\bibfnamefont {I.}~\bibnamefont {Souza}}, \bibinfo {author} {\bibfnamefont {T.}~\bibnamefont {Wilkens}}, \ and\ \bibinfo {author} {\bibfnamefont {R.~M.}\ \bibnamefont {Martin}},\ }\href {\doibase 10.1103/PhysRevB.62.1666} {\bibfield  {journal} {\bibinfo  {journal} {Physical Review B}\ }\textbf {\bibinfo {volume} {62}},\ \bibinfo {pages} {1666} (\bibinfo {year} {2000})}\BibitemShut {NoStop}\bibitem [{\citenamefont {Komissarov}\ \emph {et~al.}(2024)\citenamefont {Komissarov}, \citenamefont {Holder},\ and\ \citenamefont {Queiroz}}]{komissarovQuantumGeometricOrigin2024}  \BibitemOpen
  \bibfield  {author} {\bibinfo {author} {\bibfnamefont {I.}~\bibnamefont {Komissarov}}, \bibinfo {author} {\bibfnamefont {T.}~\bibnamefont {Holder}}, \ and\ \bibinfo {author} {\bibfnamefont {R.}~\bibnamefont {Queiroz}},\ }\href {\doibase 10.1038/s41467-024-48808-x} {\bibfield  {journal} {\bibinfo  {journal} {Nature Communications}\ }\textbf {\bibinfo {volume} {15}},\ \bibinfo {pages} {4621} (\bibinfo {year} {2024})}\BibitemShut {NoStop}\bibitem [{\citenamefont {Verma}\ and\ \citenamefont {Queiroz}(2025{\natexlab{a}})}]{vermaInstantaneous2025}  \BibitemOpen
  \bibfield  {author} {\bibinfo {author} {\bibfnamefont {N.}~\bibnamefont {Verma}}\ and\ \bibinfo {author} {\bibfnamefont {R.}~\bibnamefont {Queiroz}},\ }\href {\doibase 10.1073/pnas.2405837122} {\bibfield  {journal} {\bibinfo  {journal} {Proc. Natl. Acad. Sci. U.S.A.}\ }\textbf {\bibinfo {volume} {122}},\ \bibinfo {pages} {e2405837122} (\bibinfo {year} {2025}{\natexlab{a}})}\BibitemShut {NoStop}\bibitem [{\citenamefont {Kruchkov}\ and\ \citenamefont {Ryu}(2023)}]{kruchkov2023}  \BibitemOpen
  \bibfield  {author} {\bibinfo {author} {\bibfnamefont {A.}~\bibnamefont {Kruchkov}}\ and\ \bibinfo {author} {\bibfnamefont {S.}~\bibnamefont {Ryu}},\ }\href@noop {} {\enquote {\bibinfo {title} {Spectral sum rules reflect topological and quantum-geometric invariants},}\ } (\bibinfo {year} {2023}),\ \Eprint {http://arxiv.org/abs/2312.17318} {arxiv:2312.17318 [cond-mat]} \BibitemShut {NoStop}\bibitem [{\citenamefont {Nagaosa}\ \emph {et~al.}(2010)\citenamefont {Nagaosa}, \citenamefont {Sinova}, \citenamefont {Onoda}, \citenamefont {MacDonald},\ and\ \citenamefont {Ong}}]{nagaosaAnomalousHallEffect2010}  \BibitemOpen
  \bibfield  {author} {\bibinfo {author} {\bibfnamefont {N.}~\bibnamefont {Nagaosa}}, \bibinfo {author} {\bibfnamefont {J.}~\bibnamefont {Sinova}}, \bibinfo {author} {\bibfnamefont {S.}~\bibnamefont {Onoda}}, \bibinfo {author} {\bibfnamefont {A.~H.}\ \bibnamefont {MacDonald}}, \ and\ \bibinfo {author} {\bibfnamefont {N.~P.}\ \bibnamefont {Ong}},\ }\href {\doibase 10.1103/RevModPhys.82.1539} {\bibfield  {journal} {\bibinfo  {journal} {Reviews of Modern Physics}\ }\textbf {\bibinfo {volume} {82}},\ \bibinfo {pages} {1539} (\bibinfo {year} {2010})}\BibitemShut {NoStop}\bibitem [{\citenamefont {Thouless}\ \emph {et~al.}(1982)\citenamefont {Thouless}, \citenamefont {Kohmoto}, \citenamefont {Nightingale},\ and\ \citenamefont {{den Nijs}}}]{thoulessQuantizedHall1982}  \BibitemOpen
  \bibfield  {author} {\bibinfo {author} {\bibfnamefont {D.~J.}\ \bibnamefont {Thouless}}, \bibinfo {author} {\bibfnamefont {M.}~\bibnamefont {Kohmoto}}, \bibinfo {author} {\bibfnamefont {M.~P.}\ \bibnamefont {Nightingale}}, \ and\ \bibinfo {author} {\bibfnamefont {M.}~\bibnamefont {{den Nijs}}},\ }\href {\doibase 10.1103/PhysRevLett.49.405} {\bibfield  {journal} {\bibinfo  {journal} {Physical Review Letters}\ }\textbf {\bibinfo {volume} {49}},\ \bibinfo {pages} {405} (\bibinfo {year} {1982})}\BibitemShut {NoStop}\bibitem [{\citenamefont {Ba{\l}ut}\ \emph {et~al.}(2026)\citenamefont {Ba{\l}ut}, \citenamefont {Bradlyn}, \citenamefont {Collins},\ and\ \citenamefont {Abbamonte}}]{balutAbbamonteIXS2026}  \BibitemOpen
  \bibfield  {author} {\bibinfo {author} {\bibfnamefont {D.}~\bibnamefont {Ba{\l}ut}}, \bibinfo {author} {\bibfnamefont {B.}~\bibnamefont {Bradlyn}}, \bibinfo {author} {\bibfnamefont {M.~D.}\ \bibnamefont {Collins}}, \ and\ \bibinfo {author} {\bibfnamefont {P.}~\bibnamefont {Abbamonte}},\ }\href@noop {} {\bibfield  {journal} {\bibinfo  {journal} {arXiv}\ } (\bibinfo {year} {2026})},\ \bibinfo {note} {arXiv:2601.19054}\BibitemShut {NoStop}\bibitem [{\citenamefont {Kang}\ \emph {et~al.}(2025)\citenamefont {Kang}, \citenamefont {Kim}, \citenamefont {Qian}, \citenamefont {Neves}, \citenamefont {Ye}, \citenamefont {Jung}, \citenamefont {Puntel}, \citenamefont {Mazzola}, \citenamefont {Fang}, \citenamefont {Jozwiak}, \citenamefont {Bostwick}, \citenamefont {Rotenberg}, \citenamefont {Fuji}, \citenamefont {Vobornik}, \citenamefont {Park}, \citenamefont {Checkelsky}, \citenamefont {Yang},\ and\ \citenamefont {Comin}}]{kangMeasurementsQuantumGeometric2025}  \BibitemOpen
  \bibfield  {author} {\bibinfo {author} {\bibfnamefont {M.}~\bibnamefont {Kang}}, \bibinfo {author} {\bibfnamefont {S.}~\bibnamefont {Kim}}, \bibinfo {author} {\bibfnamefont {Y.}~\bibnamefont {Qian}}, \bibinfo {author} {\bibfnamefont {P.~M.}\ \bibnamefont {Neves}}, \bibinfo {author} {\bibfnamefont {L.}~\bibnamefont {Ye}}, \bibinfo {author} {\bibfnamefont {J.}~\bibnamefont {Jung}}, \bibinfo {author} {\bibfnamefont {D.}~\bibnamefont {Puntel}}, \bibinfo {author} {\bibfnamefont {F.}~\bibnamefont {Mazzola}}, \bibinfo {author} {\bibfnamefont {S.}~\bibnamefont {Fang}}, \bibinfo {author} {\bibfnamefont {C.}~\bibnamefont {Jozwiak}}, \bibinfo {author} {\bibfnamefont {A.}~\bibnamefont {Bostwick}}, \bibinfo {author} {\bibfnamefont {E.}~\bibnamefont {Rotenberg}}, \bibinfo {author} {\bibfnamefont {J.}~\bibnamefont {Fuji}}, \bibinfo {author} {\bibfnamefont {I.}~\bibnamefont {Vobornik}}, \bibinfo {author} {\bibfnamefont {J.-H.}\ \bibnamefont {Park}}, \bibinfo {author} {\bibfnamefont {J.~G.}\ \bibnamefont {Checkelsky}},
  \bibinfo {author} {\bibfnamefont {B.-J.}\ \bibnamefont {Yang}}, \ and\ \bibinfo {author} {\bibfnamefont {R.}~\bibnamefont {Comin}},\ }\href {\doibase 10.1038/s41567-024-02678-8} {\bibfield  {journal} {\bibinfo  {journal} {Nature Physics}\ }\textbf {\bibinfo {volume} {21}},\ \bibinfo {pages} {110} (\bibinfo {year} {2025})},\ \bibinfo {note} {arXiv:2412.17809}\BibitemShut {NoStop}\bibitem [{\citenamefont {Gianfrate}\ \emph {et~al.}(2020)\citenamefont {Gianfrate}, \citenamefont {Bleu}, \citenamefont {Dominici}, \citenamefont {Ardizzone}, \citenamefont {De~Giorgi}, \citenamefont {Ballarini}, \citenamefont {Lerario}, \citenamefont {West}, \citenamefont {Pfeiffer}, \citenamefont {Solnyshkov}, \citenamefont {Sanvitto},\ and\ \citenamefont {Malpuech}}]{gianfrateMeasurementQuantumGeometric2020}  \BibitemOpen
  \bibfield  {author} {\bibinfo {author} {\bibfnamefont {A.}~\bibnamefont {Gianfrate}}, \bibinfo {author} {\bibfnamefont {O.}~\bibnamefont {Bleu}}, \bibinfo {author} {\bibfnamefont {L.}~\bibnamefont {Dominici}}, \bibinfo {author} {\bibfnamefont {V.}~\bibnamefont {Ardizzone}}, \bibinfo {author} {\bibfnamefont {M.}~\bibnamefont {De~Giorgi}}, \bibinfo {author} {\bibfnamefont {D.}~\bibnamefont {Ballarini}}, \bibinfo {author} {\bibfnamefont {G.}~\bibnamefont {Lerario}}, \bibinfo {author} {\bibfnamefont {K.~W.}\ \bibnamefont {West}}, \bibinfo {author} {\bibfnamefont {L.~N.}\ \bibnamefont {Pfeiffer}}, \bibinfo {author} {\bibfnamefont {D.~D.}\ \bibnamefont {Solnyshkov}}, \bibinfo {author} {\bibfnamefont {D.}~\bibnamefont {Sanvitto}}, \ and\ \bibinfo {author} {\bibfnamefont {G.}~\bibnamefont {Malpuech}},\ }\href {\doibase 10.1038/s41586-020-1989-2} {\bibfield  {journal} {\bibinfo  {journal} {Nature}\ }\textbf {\bibinfo {volume} {578}},\ \bibinfo {pages} {381} (\bibinfo {year} {2020})}\BibitemShut {NoStop}\bibitem [{\citenamefont {Tan}\ \emph {et~al.}(2019)\citenamefont {Tan}, \citenamefont {Zhang}, \citenamefont {Yang}, \citenamefont {Chu}, \citenamefont {Zhu}, \citenamefont {Li}, \citenamefont {Yang}, \citenamefont {Song}, \citenamefont {Han}, \citenamefont {Li}, \citenamefont {Dong}, \citenamefont {Yu}, \citenamefont {Yan}, \citenamefont {Zhu},\ and\ \citenamefont {Yu}}]{tanQuantumMetric2019}  \BibitemOpen
  \bibfield  {author} {\bibinfo {author} {\bibfnamefont {X.}~\bibnamefont {Tan}}, \bibinfo {author} {\bibfnamefont {D.-W.}\ \bibnamefont {Zhang}}, \bibinfo {author} {\bibfnamefont {Z.}~\bibnamefont {Yang}}, \bibinfo {author} {\bibfnamefont {J.}~\bibnamefont {Chu}}, \bibinfo {author} {\bibfnamefont {Y.-Q.}\ \bibnamefont {Zhu}}, \bibinfo {author} {\bibfnamefont {D.}~\bibnamefont {Li}}, \bibinfo {author} {\bibfnamefont {X.}~\bibnamefont {Yang}}, \bibinfo {author} {\bibfnamefont {S.}~\bibnamefont {Song}}, \bibinfo {author} {\bibfnamefont {Z.}~\bibnamefont {Han}}, \bibinfo {author} {\bibfnamefont {Z.}~\bibnamefont {Li}}, \bibinfo {author} {\bibfnamefont {Y.}~\bibnamefont {Dong}}, \bibinfo {author} {\bibfnamefont {H.-F.}\ \bibnamefont {Yu}}, \bibinfo {author} {\bibfnamefont {H.}~\bibnamefont {Yan}}, \bibinfo {author} {\bibfnamefont {S.-L.}\ \bibnamefont {Zhu}}, \ and\ \bibinfo {author} {\bibfnamefont {Y.}~\bibnamefont {Yu}},\ }\href {\doibase 10.1103/PhysRevLett.122.210401} {\bibfield  {journal} {\bibinfo
  {journal} {Physical Review Letters}\ }\textbf {\bibinfo {volume} {122}},\ \bibinfo {pages} {210401} (\bibinfo {year} {2019})}\BibitemShut {NoStop}\bibitem [{\citenamefont {Verma}\ and\ \citenamefont {Queiroz}(2025{\natexlab{b}})}]{verma2025step}  \BibitemOpen
  \bibfield  {author} {\bibinfo {author} {\bibfnamefont {N.}~\bibnamefont {Verma}}\ and\ \bibinfo {author} {\bibfnamefont {R.}~\bibnamefont {Queiroz}},\ }\href {\doibase 10.1103/PhysRevLett.134.106403} {\bibfield  {journal} {\bibinfo  {journal} {Phys. Rev. Lett.}\ }\textbf {\bibinfo {volume} {134}},\ \bibinfo {pages} {106403} (\bibinfo {year} {2025}{\natexlab{b}})}\BibitemShut {NoStop}\bibitem [{\citenamefont {Phillips}(1968)}]{phillipsDielectric1968}  \BibitemOpen
  \bibfield  {author} {\bibinfo {author} {\bibfnamefont {J.~C.}\ \bibnamefont {Phillips}},\ }\href {\doibase 10.1103/PhysRevLett.20.550} {\bibfield  {journal} {\bibinfo  {journal} {Physical Review Letters}\ }\textbf {\bibinfo {volume} {20}},\ \bibinfo {pages} {550} (\bibinfo {year} {1968})}\BibitemShut {NoStop}\bibitem [{\citenamefont {{Van Vechten}}(1969)}]{vanVechtenQuantumDielectric1969}  \BibitemOpen
  \bibfield  {author} {\bibinfo {author} {\bibfnamefont {J.~A.}\ \bibnamefont {{Van Vechten}}},\ }\href {\doibase 10.1103/PhysRev.182.891} {\bibfield  {journal} {\bibinfo  {journal} {Physical Review}\ }\textbf {\bibinfo {volume} {182}},\ \bibinfo {pages} {891} (\bibinfo {year} {1969})}\BibitemShut {NoStop}\bibitem [{\citenamefont {Shportko}\ \emph {et~al.}(2008)\citenamefont {Shportko}, \citenamefont {Kremers}, \citenamefont {Woda}, \citenamefont {Lencer}, \citenamefont {Robertson},\ and\ \citenamefont {Wuttig}}]{shportkoResonantBondingCrystalline2008}  \BibitemOpen
  \bibfield  {author} {\bibinfo {author} {\bibfnamefont {K.}~\bibnamefont {Shportko}}, \bibinfo {author} {\bibfnamefont {S.}~\bibnamefont {Kremers}}, \bibinfo {author} {\bibfnamefont {M.}~\bibnamefont {Woda}}, \bibinfo {author} {\bibfnamefont {D.}~\bibnamefont {Lencer}}, \bibinfo {author} {\bibfnamefont {J.}~\bibnamefont {Robertson}}, \ and\ \bibinfo {author} {\bibfnamefont {M.}~\bibnamefont {Wuttig}},\ }\href {\doibase 10.1038/nmat2226} {\bibfield  {journal} {\bibinfo  {journal} {Nature Materials}\ }\textbf {\bibinfo {volume} {7}},\ \bibinfo {pages} {653} (\bibinfo {year} {2008})}\BibitemShut {NoStop}\bibitem [{\citenamefont {Wuttig}\ \emph {et~al.}(2018)\citenamefont {Wuttig}, \citenamefont {Deringer}, \citenamefont {Gonze}, \citenamefont {Bichara},\ and\ \citenamefont {Raty}}]{wuttigIncipientMetalsFunctional2018a}  \BibitemOpen
  \bibfield  {author} {\bibinfo {author} {\bibfnamefont {M.}~\bibnamefont {Wuttig}}, \bibinfo {author} {\bibfnamefont {V.~L.}\ \bibnamefont {Deringer}}, \bibinfo {author} {\bibfnamefont {X.}~\bibnamefont {Gonze}}, \bibinfo {author} {\bibfnamefont {C.}~\bibnamefont {Bichara}}, \ and\ \bibinfo {author} {\bibfnamefont {J.-Y.}\ \bibnamefont {Raty}},\ }\href {\doibase 10.1002/adma.201803777} {\bibfield  {journal} {\bibinfo  {journal} {Advanced Materials}\ }\textbf {\bibinfo {volume} {30}},\ \bibinfo {pages} {1803777} (\bibinfo {year} {2018})}\BibitemShut {NoStop}\bibitem [{\citenamefont {Wuttig}(2009)}]{wuttigPhaseChangeMaterials2009}  \BibitemOpen
  \bibfield  {author} {\bibinfo {author} {\bibfnamefont {M.}~\bibnamefont {Wuttig}},\ }\href {\doibase 10.1002/pssb.200982010} {\bibfield  {journal} {\bibinfo  {journal} {physica status solidi (b)}\ }\textbf {\bibinfo {volume} {246}},\ \bibinfo {pages} {1820} (\bibinfo {year} {2009})}\BibitemShut {NoStop}\bibitem [{\citenamefont {Tremel}\ and\ \citenamefont {Hoffmann}(1987)}]{tremelSquareNetsMaingroup1987}  \BibitemOpen
  \bibfield  {author} {\bibinfo {author} {\bibfnamefont {W.}~\bibnamefont {Tremel}}\ and\ \bibinfo {author} {\bibfnamefont {R.}~\bibnamefont {Hoffmann}},\ }\href {\doibase 10.1021/ja00235a021} {\bibfield  {journal} {\bibinfo  {journal} {Journal of the American Chemical Society}\ }\textbf {\bibinfo {volume} {109}},\ \bibinfo {pages} {124} (\bibinfo {year} {1987})}\BibitemShut {NoStop}\bibitem [{\citenamefont {Papoian}\ and\ \citenamefont {Hoffmann}(2000)}]{a.papoianHypervalentBondingOne2000a}  \BibitemOpen
  \bibfield  {author} {\bibinfo {author} {\bibfnamefont {G.~A.}\ \bibnamefont {Papoian}}\ and\ \bibinfo {author} {\bibfnamefont {R.}~\bibnamefont {Hoffmann}},\ }\href {\doibase 10.1002/1521-3773(20000717)39:14<2408::AID-ANIE2408>3.0.CO;2-U} {\bibfield  {journal} {\bibinfo  {journal} {Angewandte Chemie International Edition}\ }\textbf {\bibinfo {volume} {39}},\ \bibinfo {pages} {2408} (\bibinfo {year} {2000})}\BibitemShut {NoStop}\bibitem [{\citenamefont {Klemenz}\ \emph {et~al.}(2019)\citenamefont {Klemenz}, \citenamefont {Lei},\ and\ \citenamefont {Schoop}}]{klemenzTopologicalSemimetalsSquareNet2019}  \BibitemOpen
  \bibfield  {author} {\bibinfo {author} {\bibfnamefont {S.}~\bibnamefont {Klemenz}}, \bibinfo {author} {\bibfnamefont {S.}~\bibnamefont {Lei}}, \ and\ \bibinfo {author} {\bibfnamefont {L.~M.}\ \bibnamefont {Schoop}},\ }\href {\doibase 10.1146/annurev-matsci-070218-010114} {\bibfield  {journal} {\bibinfo  {journal} {Annual Review of Materials Research}\ }\textbf {\bibinfo {volume} {49}},\ \bibinfo {pages} {185} (\bibinfo {year} {2019})}\BibitemShut {NoStop}\bibitem [{\citenamefont {Klemenz}\ \emph {et~al.}(2020)\citenamefont {Klemenz}, \citenamefont {Hay}, \citenamefont {Teicher}, \citenamefont {Topp}, \citenamefont {Cano},\ and\ \citenamefont {Schoop}}]{klemenzRoleDelocalizedChemical2020a}  \BibitemOpen
  \bibfield  {author} {\bibinfo {author} {\bibfnamefont {S.}~\bibnamefont {Klemenz}}, \bibinfo {author} {\bibfnamefont {A.~K.}\ \bibnamefont {Hay}}, \bibinfo {author} {\bibfnamefont {S.~M.~L.}\ \bibnamefont {Teicher}}, \bibinfo {author} {\bibfnamefont {A.}~\bibnamefont {Topp}}, \bibinfo {author} {\bibfnamefont {J.}~\bibnamefont {Cano}}, \ and\ \bibinfo {author} {\bibfnamefont {L.~M.}\ \bibnamefont {Schoop}},\ }\href {\doibase 10.1021/jacs.0c01227} {\bibfield  {journal} {\bibinfo  {journal} {Journal of the American Chemical Society}\ }\textbf {\bibinfo {volume} {142}},\ \bibinfo {pages} {6350} (\bibinfo {year} {2020})}\BibitemShut {NoStop}\bibitem [{\citenamefont {Lee}\ and\ \citenamefont {Elliott}(2020)}]{leeElliottHyperbonding2020}  \BibitemOpen
  \bibfield  {author} {\bibinfo {author} {\bibfnamefont {T.~H.}\ \bibnamefont {Lee}}\ and\ \bibinfo {author} {\bibfnamefont {S.~R.}\ \bibnamefont {Elliott}},\ }\href {\doibase 10.1002/adma.202000340} {\bibfield  {journal} {\bibinfo  {journal} {Advanced Materials}\ }\textbf {\bibinfo {volume} {32}},\ \bibinfo {pages} {2000340} (\bibinfo {year} {2020})}\BibitemShut {NoStop}\bibitem [{\citenamefont {Manj{\'o}n}\ \emph {et~al.}(2024)\citenamefont {Manj{\'o}n}, \citenamefont {Osman}, \citenamefont {Savastano},\ and\ \citenamefont {Vegas}}]{manjonherreraElectrondeficientMulticenterBonding2024}  \BibitemOpen
  \bibfield  {author} {\bibinfo {author} {\bibfnamefont {F.~J.}\ \bibnamefont {Manj{\'o}n}}, \bibinfo {author} {\bibfnamefont {H.~H.}\ \bibnamefont {Osman}}, \bibinfo {author} {\bibfnamefont {M.}~\bibnamefont {Savastano}}, \ and\ \bibinfo {author} {\bibfnamefont {{\'A}.}~\bibnamefont {Vegas}},\ }\href {\doibase 10.3390/ma17122840} {\bibfield  {journal} {\bibinfo  {journal} {Materials}\ }\textbf {\bibinfo {volume} {17}},\ \bibinfo {pages} {2840} (\bibinfo {year} {2024})}\BibitemShut {NoStop}\bibitem [{\citenamefont {Kim}\ and\ \citenamefont {{Head-Gordon}}(2024)}]{kimEquivalencePolarizabilityBond2024a}  \BibitemOpen
  \bibfield  {author} {\bibinfo {author} {\bibfnamefont {L.}~\bibnamefont {Kim}}\ and\ \bibinfo {author} {\bibfnamefont {T.}~\bibnamefont {{Head-Gordon}}},\ }\href {\doibase 10.1039/D4CP03337J} {\bibfield  {journal} {\bibinfo  {journal} {Physical Chemistry Chemical Physics}\ }\textbf {\bibinfo {volume} {26}},\ \bibinfo {pages} {27459} (\bibinfo {year} {2024})}\BibitemShut {NoStop}\bibitem [{\citenamefont {Giambiagi}\ \emph {et~al.}(2000)\citenamefont {Giambiagi}, \citenamefont {de~Giambiagi}, \citenamefont {Silva},\ and\ \citenamefont {de~Figueiredo}}]{giambiagiMulticenterBondIndices2000}  \BibitemOpen
  \bibfield  {author} {\bibinfo {author} {\bibfnamefont {M.}~\bibnamefont {Giambiagi}}, \bibinfo {author} {\bibfnamefont {M.~S.}\ \bibnamefont {de~Giambiagi}}, \bibinfo {author} {\bibfnamefont {C.~D. d.~S.}\ \bibnamefont {Silva}}, \ and\ \bibinfo {author} {\bibfnamefont {A.~P.}\ \bibnamefont {de~Figueiredo}},\ }\href {\doibase 10.1039/B002009P} {\bibfield  {journal} {\bibinfo  {journal} {Physical Chemistry Chemical Physics}\ }\textbf {\bibinfo {volume} {2}},\ \bibinfo {pages} {3381} (\bibinfo {year} {2000})}\BibitemShut {NoStop}\bibitem [{\citenamefont {Bultinck}\ \emph {et~al.}(2005)\citenamefont {Bultinck}, \citenamefont {Ponec},\ and\ \citenamefont {Van~Damme}}]{bultinckMulticenterBond2005}  \BibitemOpen
  \bibfield  {author} {\bibinfo {author} {\bibfnamefont {P.}~\bibnamefont {Bultinck}}, \bibinfo {author} {\bibfnamefont {R.}~\bibnamefont {Ponec}}, \ and\ \bibinfo {author} {\bibfnamefont {S.}~\bibnamefont {Van~Damme}},\ }\href {\doibase 10.1002/poc.922} {\bibfield  {journal} {\bibinfo  {journal} {Journal of Physical Organic Chemistry}\ }\textbf {\bibinfo {volume} {18}},\ \bibinfo {pages} {706} (\bibinfo {year} {2005})}\BibitemShut {NoStop}\bibitem [{\citenamefont {Pimentel}(1951)}]{pimentelTrihalide1951}  \BibitemOpen
  \bibfield  {author} {\bibinfo {author} {\bibfnamefont {G.~C.}\ \bibnamefont {Pimentel}},\ }\href {\doibase 10.1063/1.1748245} {\bibfield  {journal} {\bibinfo  {journal} {The Journal of Chemical Physics}\ }\textbf {\bibinfo {volume} {19}},\ \bibinfo {pages} {446} (\bibinfo {year} {1951})}\BibitemShut {NoStop}\bibitem [{\citenamefont {Haldane}(1988)}]{haldaneModelQuantumHall1988}  \BibitemOpen
  \bibfield  {author} {\bibinfo {author} {\bibfnamefont {F.~D.~M.}\ \bibnamefont {Haldane}},\ }\href {\doibase 10.1103/PhysRevLett.61.2015} {\bibfield  {journal} {\bibinfo  {journal} {Physical Review Letters}\ }\textbf {\bibinfo {volume} {61}},\ \bibinfo {pages} {2015} (\bibinfo {year} {1988})}\BibitemShut {NoStop}\bibitem [{\citenamefont {Coulson}(1939)}]{coulsonFractionalBondOrder1939}  \BibitemOpen
  \bibfield  {author} {\bibinfo {author} {\bibfnamefont {C.~A.}\ \bibnamefont {Coulson}},\ }\href {\doibase 10.1098/rspa.1939.0006} {\bibfield  {journal} {\bibinfo  {journal} {Proceedings of the Royal Society of London. Series A. Mathematical and Physical Sciences}\ }\textbf {\bibinfo {volume} {169}},\ \bibinfo {pages} {413} (\bibinfo {year} {1939})}\BibitemShut {NoStop}\bibitem [{\citenamefont {Sannigrahi}\ and\ \citenamefont {Kar}(1988)}]{sannigrahiThreeCentreBond1988}  \BibitemOpen
  \bibfield  {author} {\bibinfo {author} {\bibfnamefont {A.~B.}\ \bibnamefont {Sannigrahi}}\ and\ \bibinfo {author} {\bibfnamefont {T.}~\bibnamefont {Kar}},\ }\href@noop {} {\bibfield  {journal} {\bibinfo  {journal} {Chemical Physics Letters}\ }\textbf {\bibinfo {volume} {150}},\ \bibinfo {pages} {173} (\bibinfo {year} {1988})}\BibitemShut {NoStop}\bibitem [{\citenamefont {Giambiagi}\ \emph {et~al.}(1990)\citenamefont {Giambiagi}, \citenamefont {{de Giambiagi}},\ and\ \citenamefont {Mundim}}]{giambiagiDefinitionMulticenterBond1990}  \BibitemOpen
  \bibfield  {author} {\bibinfo {author} {\bibfnamefont {M.}~\bibnamefont {Giambiagi}}, \bibinfo {author} {\bibfnamefont {M.~S.}\ \bibnamefont {{de Giambiagi}}}, \ and\ \bibinfo {author} {\bibfnamefont {K.~C.}\ \bibnamefont {Mundim}},\ }\href {\doibase 10.1007/BF00671228} {\bibfield  {journal} {\bibinfo  {journal} {Structural Chemistry}\ }\textbf {\bibinfo {volume} {1}},\ \bibinfo {pages} {423} (\bibinfo {year} {1990})}\BibitemShut {NoStop}\bibitem [{\citenamefont {Graf}\ and\ \citenamefont {Vogl}(1995)}]{grafVoglElectromagnetic1995}  \BibitemOpen
  \bibfield  {author} {\bibinfo {author} {\bibfnamefont {M.}~\bibnamefont {Graf}}\ and\ \bibinfo {author} {\bibfnamefont {P.}~\bibnamefont {Vogl}},\ }\href {\doibase 10.1103/PhysRevB.51.4940} {\bibfield  {journal} {\bibinfo  {journal} {Physical Review B}\ }\textbf {\bibinfo {volume} {51}},\ \bibinfo {pages} {4940} (\bibinfo {year} {1995})}\BibitemShut {NoStop}\bibitem [{\citenamefont {Marzari}\ and\ \citenamefont {Vanderbilt}(1997)}]{marzariMaximallyLocalizedGeneralized1997}  \BibitemOpen
  \bibfield  {author} {\bibinfo {author} {\bibfnamefont {N.}~\bibnamefont {Marzari}}\ and\ \bibinfo {author} {\bibfnamefont {D.}~\bibnamefont {Vanderbilt}},\ }\href {\doibase 10.1103/PhysRevB.56.12847} {\bibfield  {journal} {\bibinfo  {journal} {Physical Review B}\ }\textbf {\bibinfo {volume} {56}},\ \bibinfo {pages} {12847} (\bibinfo {year} {1997})}\BibitemShut {NoStop}\bibitem [{\citenamefont {Bradlyn}\ \emph {et~al.}(2017)\citenamefont {Bradlyn}, \citenamefont {Elcoro}, \citenamefont {Cano}, \citenamefont {Vergniory}, \citenamefont {Wang}, \citenamefont {Felser}, \citenamefont {Aroyo},\ and\ \citenamefont {Bernevig}}]{bradlynTQC2017}  \BibitemOpen
  \bibfield  {author} {\bibinfo {author} {\bibfnamefont {B.}~\bibnamefont {Bradlyn}}, \bibinfo {author} {\bibfnamefont {L.}~\bibnamefont {Elcoro}}, \bibinfo {author} {\bibfnamefont {J.}~\bibnamefont {Cano}}, \bibinfo {author} {\bibfnamefont {M.~G.}\ \bibnamefont {Vergniory}}, \bibinfo {author} {\bibfnamefont {Z.}~\bibnamefont {Wang}}, \bibinfo {author} {\bibfnamefont {C.}~\bibnamefont {Felser}}, \bibinfo {author} {\bibfnamefont {M.~I.}\ \bibnamefont {Aroyo}}, \ and\ \bibinfo {author} {\bibfnamefont {B.~A.}\ \bibnamefont {Bernevig}},\ }\href {\doibase 10.1038/nature23268} {\bibfield  {journal} {\bibinfo  {journal} {Nature}\ }\textbf {\bibinfo {volume} {547}},\ \bibinfo {pages} {298} (\bibinfo {year} {2017})}\BibitemShut {NoStop}\bibitem [{\citenamefont {H{\"u}ckel}(1931)}]{huckelBenzolproblem1931}  \BibitemOpen
  \bibfield  {author} {\bibinfo {author} {\bibfnamefont {E.}~\bibnamefont {H{\"u}ckel}},\ }\href {\doibase 10.1007/BF01339530} {\bibfield  {journal} {\bibinfo  {journal} {Zeitschrift f{\"u}r Physik}\ }\textbf {\bibinfo {volume} {70}},\ \bibinfo {pages} {204} (\bibinfo {year} {1931})}\BibitemShut {NoStop}\bibitem [{\citenamefont {Kubo}(1966)}]{kuboFluctuationdissipationTheorem}  \BibitemOpen
  \bibfield  {author} {\bibinfo {author} {\bibfnamefont {R.}~\bibnamefont {Kubo}},\ }\href {\doibase 10.1088/0034-4885/29/1/306} {\bibfield  {journal} {\bibinfo  {journal} {Reports on Progress in Physics}\ }\textbf {\bibinfo {volume} {29}},\ \bibinfo {pages} {255} (\bibinfo {year} {1966})}\BibitemShut {NoStop}\bibitem [{\citenamefont {Souza}\ \emph {et~al.}(2025)\citenamefont {Souza}, \citenamefont {Martin},\ and\ \citenamefont {Stengel}}]{Souza2025}  \BibitemOpen
  \bibfield  {author} {\bibinfo {author} {\bibfnamefont {I.}~\bibnamefont {Souza}}, \bibinfo {author} {\bibfnamefont {R.~M.}\ \bibnamefont {Martin}}, \ and\ \bibinfo {author} {\bibfnamefont {M.}~\bibnamefont {Stengel}},\ }\href {\doibase 10.21468/SciPostPhys.18.4.127} {\bibfield  {journal} {\bibinfo  {journal} {SciPost Physics}\ }\textbf {\bibinfo {volume} {18}} (\bibinfo {year} {2025}),\ 10.21468/SciPostPhys.18.4.127}\BibitemShut {NoStop}\bibitem [{\citenamefont {Verma}\ and\ \citenamefont {Queiroz}(2025{\natexlab{c}})}]{vermaFrameworkMeasureQuantum2025}  \BibitemOpen
  \bibfield  {author} {\bibinfo {author} {\bibfnamefont {N.}~\bibnamefont {Verma}}\ and\ \bibinfo {author} {\bibfnamefont {R.}~\bibnamefont {Queiroz}},\ }\href {\doibase 10.1103/PhysRevLett.134.106403} {\bibfield  {journal} {\bibinfo  {journal} {Physical Review Letters}\ }\textbf {\bibinfo {volume} {134}},\ \bibinfo {pages} {106403} (\bibinfo {year} {2025}{\natexlab{c}})}\BibitemShut {NoStop}\bibitem [{\citenamefont {Souza}\ and\ \citenamefont {Vanderbilt}(2008)}]{souzaDichroic$f$sumRule2008}  \BibitemOpen
  \bibfield  {author} {\bibinfo {author} {\bibfnamefont {I.}~\bibnamefont {Souza}}\ and\ \bibinfo {author} {\bibfnamefont {D.}~\bibnamefont {Vanderbilt}},\ }\href {\doibase 10.1103/PhysRevB.77.054438} {\bibfield  {journal} {\bibinfo  {journal} {Physical Review B}\ }\textbf {\bibinfo {volume} {77}},\ \bibinfo {pages} {054438} (\bibinfo {year} {2008})}\BibitemShut {NoStop}\bibitem [{\citenamefont {Resta}(2020)}]{restaXrayCircularDichroism2020}  \BibitemOpen
  \bibfield  {author} {\bibinfo {author} {\bibfnamefont {R.}~\bibnamefont {Resta}},\ }\href {\doibase 10.1103/PhysRevResearch.2.023139} {\bibfield  {journal} {\bibinfo  {journal} {Physical Review Research}\ }\textbf {\bibinfo {volume} {2}},\ \bibinfo {pages} {023139} (\bibinfo {year} {2020})},\ \Eprint {http://arxiv.org/abs/2003.04256} {arXiv:2003.04256 [cond-mat]} \BibitemShut {NoStop}\bibitem [{\citenamefont {London}(1937)}]{london1937}  \BibitemOpen
  \bibfield  {author} {\bibinfo {author} {\bibfnamefont {F.}~\bibnamefont {London}},\ }\href {\doibase 10.1051/jphysrad:01937008010039700} {\bibfield  {journal} {\bibinfo  {journal} {Journal de Physique et le Radium}\ }\textbf {\bibinfo {volume} {8}},\ \bibinfo {pages} {397} (\bibinfo {year} {1937})}\BibitemShut {NoStop}\bibitem [{\citenamefont {Pople}(1958)}]{pople1958}  \BibitemOpen
  \bibfield  {author} {\bibinfo {author} {\bibfnamefont {J.~A.}\ \bibnamefont {Pople}},\ }\href {\doibase 10.1080/00268975800100211} {\bibfield  {journal} {\bibinfo  {journal} {Molecular Physics}\ }\textbf {\bibinfo {volume} {1}},\ \bibinfo {pages} {175} (\bibinfo {year} {1958})}\BibitemShut {NoStop}\bibitem [{\citenamefont {Ceresoli}\ \emph {et~al.}(2006)\citenamefont {Ceresoli}, \citenamefont {Thonhauser}, \citenamefont {Vanderbilt},\ and\ \citenamefont {Resta}}]{ceresoliOrbitalMagnetization2006}  \BibitemOpen
  \bibfield  {author} {\bibinfo {author} {\bibfnamefont {D.}~\bibnamefont {Ceresoli}}, \bibinfo {author} {\bibfnamefont {T.}~\bibnamefont {Thonhauser}}, \bibinfo {author} {\bibfnamefont {D.}~\bibnamefont {Vanderbilt}}, \ and\ \bibinfo {author} {\bibfnamefont {R.}~\bibnamefont {Resta}},\ }\href {\doibase 10.1103/PhysRevB.74.024408} {\bibfield  {journal} {\bibinfo  {journal} {Physical Review B}\ }\textbf {\bibinfo {volume} {74}},\ \bibinfo {pages} {024408} (\bibinfo {year} {2006})}\BibitemShut {NoStop}\bibitem [{\citenamefont {Bianco}\ and\ \citenamefont {Resta}(2013)}]{bianco2013}  \BibitemOpen
  \bibfield  {author} {\bibinfo {author} {\bibfnamefont {R.}~\bibnamefont {Bianco}}\ and\ \bibinfo {author} {\bibfnamefont {R.}~\bibnamefont {Resta}},\ }\href {\doibase 10.1103/PhysRevLett.110.087202} {\bibfield  {journal} {\bibinfo  {journal} {Phys. Rev. Lett.}\ }\textbf {\bibinfo {volume} {110}},\ \bibinfo {pages} {087202} (\bibinfo {year} {2013})}\BibitemShut {NoStop}\bibitem [{\citenamefont {Pauling}(1936)}]{paulingDiamagneticAnisotropy1936}  \BibitemOpen
  \bibfield  {author} {\bibinfo {author} {\bibfnamefont {L.}~\bibnamefont {Pauling}},\ }\href {\doibase 10.1063/1.1749766} {\bibfield  {journal} {\bibinfo  {journal} {The Journal of Chemical Physics}\ }\textbf {\bibinfo {volume} {4}},\ \bibinfo {pages} {673} (\bibinfo {year} {1936})}\BibitemShut {NoStop}\bibitem [{\citenamefont {Pople}(1956)}]{popleRingCurrents1956}  \BibitemOpen
  \bibfield  {author} {\bibinfo {author} {\bibfnamefont {J.~A.}\ \bibnamefont {Pople}},\ }\href {\doibase 10.1063/1.1742701} {\bibfield  {journal} {\bibinfo  {journal} {The Journal of Chemical Physics}\ }\textbf {\bibinfo {volume} {24}},\ \bibinfo {pages} {1111} (\bibinfo {year} {1956})}\BibitemShut {NoStop}\bibitem [{\citenamefont {Schleyer}\ \emph {et~al.}(1996)\citenamefont {Schleyer}, \citenamefont {Maerker}, \citenamefont {Dransfeld}, \citenamefont {Jiao},\ and\ \citenamefont {van Eikema~Hommes}}]{schleyerNICS1996}  \BibitemOpen
  \bibfield  {author} {\bibinfo {author} {\bibfnamefont {P.~v.~R.}\ \bibnamefont {Schleyer}}, \bibinfo {author} {\bibfnamefont {C.}~\bibnamefont {Maerker}}, \bibinfo {author} {\bibfnamefont {A.}~\bibnamefont {Dransfeld}}, \bibinfo {author} {\bibfnamefont {H.}~\bibnamefont {Jiao}}, \ and\ \bibinfo {author} {\bibfnamefont {N.~J.~R.}\ \bibnamefont {van Eikema~Hommes}},\ }\href {\doibase 10.1021/ja960582d} {\bibfield  {journal} {\bibinfo  {journal} {Journal of the American Chemical Society}\ }\textbf {\bibinfo {volume} {118}},\ \bibinfo {pages} {6317} (\bibinfo {year} {1996})}\BibitemShut {NoStop}\bibitem [{\citenamefont {Gershoni-Poranne}\ and\ \citenamefont {Stanger}(2015)}]{gershoniporanneNICS2015}  \BibitemOpen
  \bibfield  {author} {\bibinfo {author} {\bibfnamefont {R.}~\bibnamefont {Gershoni-Poranne}}\ and\ \bibinfo {author} {\bibfnamefont {A.}~\bibnamefont {Stanger}},\ }\href {\doibase 10.1039/C5CS00114E} {\bibfield  {journal} {\bibinfo  {journal} {Chemical Society Reviews}\ }\textbf {\bibinfo {volume} {44}},\ \bibinfo {pages} {6597} (\bibinfo {year} {2015})}\BibitemShut {NoStop}\bibitem [{\citenamefont {Bianco}\ and\ \citenamefont {Resta}(2011)}]{biancoMappingTopologicalOrder2011}  \BibitemOpen
  \bibfield  {author} {\bibinfo {author} {\bibfnamefont {R.}~\bibnamefont {Bianco}}\ and\ \bibinfo {author} {\bibfnamefont {R.}~\bibnamefont {Resta}},\ }\href {\doibase 10.1103/PhysRevB.84.241106} {\bibfield  {journal} {\bibinfo  {journal} {Phys. Rev. B}\ }\textbf {\bibinfo {volume} {84}},\ \bibinfo {pages} {241106} (\bibinfo {year} {2011})}\BibitemShut {NoStop}\bibitem [{\citenamefont {Fukui}\ \emph {et~al.}(2005)\citenamefont {Fukui}, \citenamefont {Hatsugai},\ and\ \citenamefont {Suzuki}}]{fukuiChernNumbers2005}  \BibitemOpen
  \bibfield  {author} {\bibinfo {author} {\bibfnamefont {T.}~\bibnamefont {Fukui}}, \bibinfo {author} {\bibfnamefont {Y.}~\bibnamefont {Hatsugai}}, \ and\ \bibinfo {author} {\bibfnamefont {H.}~\bibnamefont {Suzuki}},\ }\href {\doibase 10.1143/JPSJ.74.1674} {\bibfield  {journal} {\bibinfo  {journal} {Journal of the Physical Society of Japan}\ }\textbf {\bibinfo {volume} {74}},\ \bibinfo {pages} {1674} (\bibinfo {year} {2005})}\BibitemShut {NoStop}\bibitem [{\citenamefont {Otrokov}\ \emph {et~al.}(2019)\citenamefont {Otrokov}, \citenamefont {Rusinov}, \citenamefont {Blanco-Rey}, \citenamefont {Hoffmann}, \citenamefont {Vyazovskaya}, \citenamefont {Eremeev}, \citenamefont {Ernst}, \citenamefont {Echenique}, \citenamefont {Arnau},\ and\ \citenamefont {Chulkov}}]{otrokovUniqueThicknessDependent2019}  \BibitemOpen
  \bibfield  {author} {\bibinfo {author} {\bibfnamefont {M.~M.}\ \bibnamefont {Otrokov}}, \bibinfo {author} {\bibfnamefont {I.~P.}\ \bibnamefont {Rusinov}}, \bibinfo {author} {\bibfnamefont {M.}~\bibnamefont {Blanco-Rey}}, \bibinfo {author} {\bibfnamefont {M.}~\bibnamefont {Hoffmann}}, \bibinfo {author} {\bibfnamefont {A.~Y.}\ \bibnamefont {Vyazovskaya}}, \bibinfo {author} {\bibfnamefont {S.~V.}\ \bibnamefont {Eremeev}}, \bibinfo {author} {\bibfnamefont {A.}~\bibnamefont {Ernst}}, \bibinfo {author} {\bibfnamefont {P.~M.}\ \bibnamefont {Echenique}}, \bibinfo {author} {\bibfnamefont {A.}~\bibnamefont {Arnau}}, \ and\ \bibinfo {author} {\bibfnamefont {E.~V.}\ \bibnamefont {Chulkov}},\ }\href {\doibase 10.1103/PhysRevLett.122.107202} {\bibfield  {journal} {\bibinfo  {journal} {Physical Review Letters}\ }\textbf {\bibinfo {volume} {122}},\ \bibinfo {pages} {107202} (\bibinfo {year} {2019})}\BibitemShut {NoStop}\bibitem [{\citenamefont {Deng}\ \emph {et~al.}(2020)\citenamefont {Deng}, \citenamefont {Yu}, \citenamefont {Shi}, \citenamefont {Guo}, \citenamefont {Xu}, \citenamefont {Wang}, \citenamefont {Chen},\ and\ \citenamefont {Zhang}}]{dengQuantumAnomalousHall2020}  \BibitemOpen
  \bibfield  {author} {\bibinfo {author} {\bibfnamefont {Y.}~\bibnamefont {Deng}}, \bibinfo {author} {\bibfnamefont {Y.}~\bibnamefont {Yu}}, \bibinfo {author} {\bibfnamefont {M.~Z.}\ \bibnamefont {Shi}}, \bibinfo {author} {\bibfnamefont {Z.}~\bibnamefont {Guo}}, \bibinfo {author} {\bibfnamefont {Z.}~\bibnamefont {Xu}}, \bibinfo {author} {\bibfnamefont {J.}~\bibnamefont {Wang}}, \bibinfo {author} {\bibfnamefont {X.~H.}\ \bibnamefont {Chen}}, \ and\ \bibinfo {author} {\bibfnamefont {Y.}~\bibnamefont {Zhang}},\ }\href {\doibase 10.1126/science.aax8156} {\bibfield  {journal} {\bibinfo  {journal} {Science}\ }\textbf {\bibinfo {volume} {367}},\ \bibinfo {pages} {895} (\bibinfo {year} {2020})}\BibitemShut {NoStop}\bibitem [{\citenamefont {Hoffmann}(1987)}]{hoffmannHowChemistryPhysics1987}  \BibitemOpen
  \bibfield  {author} {\bibinfo {author} {\bibfnamefont {R.}~\bibnamefont {Hoffmann}},\ }\href {\doibase 10.1002/anie.198708461} {\bibfield  {journal} {\bibinfo  {journal} {Angewandte Chemie International Edition in English}\ }\textbf {\bibinfo {volume} {26}},\ \bibinfo {pages} {846} (\bibinfo {year} {1987})}\BibitemShut {NoStop}\bibitem [{\citenamefont {Hasan}\ and\ \citenamefont {Kane}(2010)}]{hasanColloquiumTopologicalInsulators2010a}  \BibitemOpen
  \bibfield  {author} {\bibinfo {author} {\bibfnamefont {M.~Z.}\ \bibnamefont {Hasan}}\ and\ \bibinfo {author} {\bibfnamefont {C.~L.}\ \bibnamefont {Kane}},\ }\href {\doibase 10.1103/RevModPhys.82.3045} {\bibfield  {journal} {\bibinfo  {journal} {Reviews of Modern Physics}\ }\textbf {\bibinfo {volume} {82}},\ \bibinfo {pages} {3045} (\bibinfo {year} {2010})}\BibitemShut {NoStop}\bibitem [{\citenamefont {Geng}\ \emph {et~al.}(2025)\citenamefont {Geng}, \citenamefont {Liu}, \citenamefont {Xu}, \citenamefont {Tang}, \citenamefont {Kawazoe},\ and\ \citenamefont {Wang}}]{gengEmpoweringMaterialsScience2025}  \BibitemOpen
  \bibfield  {author} {\bibinfo {author} {\bibfnamefont {W.-T.}\ \bibnamefont {Geng}}, \bibinfo {author} {\bibfnamefont {Y.-C.}\ \bibnamefont {Liu}}, \bibinfo {author} {\bibfnamefont {N.}~\bibnamefont {Xu}}, \bibinfo {author} {\bibfnamefont {G.}~\bibnamefont {Tang}}, \bibinfo {author} {\bibfnamefont {Y.}~\bibnamefont {Kawazoe}}, \ and\ \bibinfo {author} {\bibfnamefont {V.}~\bibnamefont {Wang}},\ }\href {\doibase 10.1038/s41596-025-01160-w} {\bibfield  {journal} {\bibinfo  {journal} {Nature Protocols}\ }\textbf {\bibinfo {volume} {20}},\ \bibinfo {pages} {3143} (\bibinfo {year} {2025})}\BibitemShut {NoStop}\bibitem [{\citenamefont {Pizzi}\ \emph {et~al.}(2020)\citenamefont {Pizzi}, \citenamefont {Vitale}, \citenamefont {Arita}, \citenamefont {Bl{\"u}gel}, \citenamefont {Freimuth}, \citenamefont {G{\'e}ranton}, \citenamefont {Gibertini}, \citenamefont {Gresch}, \citenamefont {Johnson}, \citenamefont {Koretsune}, \citenamefont {{Iba{\~n}ez-Azpiroz}}, \citenamefont {Lee}, \citenamefont {Lihm}, \citenamefont {Marchand}, \citenamefont {Marrazzo}, \citenamefont {Mokrousov}, \citenamefont {Mustafa}, \citenamefont {Nohara}, \citenamefont {Nomura}, \citenamefont {Paulatto}, \citenamefont {Ponc{\'e}}, \citenamefont {Ponweiser}, \citenamefont {Qiao}, \citenamefont {Th{\"o}le}, \citenamefont {Tsirkin}, \citenamefont {Wierzbowska}, \citenamefont {Marzari}, \citenamefont {Vanderbilt}, \citenamefont {Souza}, \citenamefont {Mostofi},\ and\ \citenamefont {Yates}}]{pizziWannier90CommunityCode2020}  \BibitemOpen
  \bibfield  {author} {\bibinfo {author} {\bibfnamefont {G.}~\bibnamefont {Pizzi}}, \bibinfo {author} {\bibfnamefont {V.}~\bibnamefont {Vitale}}, \bibinfo {author} {\bibfnamefont {R.}~\bibnamefont {Arita}}, \bibinfo {author} {\bibfnamefont {S.}~\bibnamefont {Bl{\"u}gel}}, \bibinfo {author} {\bibfnamefont {F.}~\bibnamefont {Freimuth}}, \bibinfo {author} {\bibfnamefont {G.}~\bibnamefont {G{\'e}ranton}}, \bibinfo {author} {\bibfnamefont {M.}~\bibnamefont {Gibertini}}, \bibinfo {author} {\bibfnamefont {D.}~\bibnamefont {Gresch}}, \bibinfo {author} {\bibfnamefont {C.}~\bibnamefont {Johnson}}, \bibinfo {author} {\bibfnamefont {T.}~\bibnamefont {Koretsune}}, \bibinfo {author} {\bibfnamefont {J.}~\bibnamefont {{Iba{\~n}ez-Azpiroz}}}, \bibinfo {author} {\bibfnamefont {H.}~\bibnamefont {Lee}}, \bibinfo {author} {\bibfnamefont {J.-M.}\ \bibnamefont {Lihm}}, \bibinfo {author} {\bibfnamefont {D.}~\bibnamefont {Marchand}}, \bibinfo {author} {\bibfnamefont {A.}~\bibnamefont {Marrazzo}}, \bibinfo {author} {\bibfnamefont
  {Y.}~\bibnamefont {Mokrousov}}, \bibinfo {author} {\bibfnamefont {J.~I.}\ \bibnamefont {Mustafa}}, \bibinfo {author} {\bibfnamefont {Y.}~\bibnamefont {Nohara}}, \bibinfo {author} {\bibfnamefont {Y.}~\bibnamefont {Nomura}}, \bibinfo {author} {\bibfnamefont {L.}~\bibnamefont {Paulatto}}, \bibinfo {author} {\bibfnamefont {S.}~\bibnamefont {Ponc{\'e}}}, \bibinfo {author} {\bibfnamefont {T.}~\bibnamefont {Ponweiser}}, \bibinfo {author} {\bibfnamefont {J.}~\bibnamefont {Qiao}}, \bibinfo {author} {\bibfnamefont {F.}~\bibnamefont {Th{\"o}le}}, \bibinfo {author} {\bibfnamefont {S.~S.}\ \bibnamefont {Tsirkin}}, \bibinfo {author} {\bibfnamefont {M.}~\bibnamefont {Wierzbowska}}, \bibinfo {author} {\bibfnamefont {N.}~\bibnamefont {Marzari}}, \bibinfo {author} {\bibfnamefont {D.}~\bibnamefont {Vanderbilt}}, \bibinfo {author} {\bibfnamefont {I.}~\bibnamefont {Souza}}, \bibinfo {author} {\bibfnamefont {A.~A.}\ \bibnamefont {Mostofi}}, \ and\ \bibinfo {author} {\bibfnamefont {J.~R.}\ \bibnamefont {Yates}},\ }\href {\doibase
  10.1088/1361-648X/ab51ff} {\bibfield  {journal} {\bibinfo  {journal} {Journal of Physics: Condensed Matter}\ }\textbf {\bibinfo {volume} {32}},\ \bibinfo {pages} {165902} (\bibinfo {year} {2020})}\BibitemShut {NoStop}\bibitem [{\citenamefont {Souza}\ \emph {et~al.}(2001)\citenamefont {Souza}, \citenamefont {Marzari},\ and\ \citenamefont {Vanderbilt}}]{souzaMaximallyLocalizedEntangled2001}  \BibitemOpen
  \bibfield  {author} {\bibinfo {author} {\bibfnamefont {I.}~\bibnamefont {Souza}}, \bibinfo {author} {\bibfnamefont {N.}~\bibnamefont {Marzari}}, \ and\ \bibinfo {author} {\bibfnamefont {D.}~\bibnamefont {Vanderbilt}},\ }\href {\doibase 10.1103/PhysRevB.65.035109} {\bibfield  {journal} {\bibinfo  {journal} {Physical Review B}\ }\textbf {\bibinfo {volume} {65}},\ \bibinfo {pages} {035109} (\bibinfo {year} {2001})}\BibitemShut {NoStop}\bibitem [{\citenamefont {Maintz}\ \emph {et~al.}(2016)\citenamefont {Maintz}, \citenamefont {Deringer}, \citenamefont {Tchougr{\'e}eff},\ and\ \citenamefont {Dronskowski}}]{maintzLOBSTERToolExtract2016}  \BibitemOpen
  \bibfield  {author} {\bibinfo {author} {\bibfnamefont {S.}~\bibnamefont {Maintz}}, \bibinfo {author} {\bibfnamefont {V.~L.}\ \bibnamefont {Deringer}}, \bibinfo {author} {\bibfnamefont {A.~L.}\ \bibnamefont {Tchougr{\'e}eff}}, \ and\ \bibinfo {author} {\bibfnamefont {R.}~\bibnamefont {Dronskowski}},\ }\href {\doibase 10.1002/jcc.24300} {\bibfield  {journal} {\bibinfo  {journal} {Journal of Computational Chemistry}\ }\textbf {\bibinfo {volume} {37}},\ \bibinfo {pages} {1030} (\bibinfo {year} {2016})}\BibitemShut {NoStop}\end{thebibliography}

\clearpage
\onecolumngrid
\renewcommand{\thefigure}{S\arabic{figure}}
\renewcommand{\thetable}{S\arabic{table}}
\renewcommand{\theequation}{S\arabic{equation}}
\ifdefined\theHequation\renewcommand{\theHequation}{S\arabic{equation}}\fi
\ifdefined\theHfigure\renewcommand{\theHfigure}{S\arabic{figure}}\fi
\ifdefined\theHtable\renewcommand{\theHtable}{S\arabic{table}}\fi
\setcounter{figure}{0}
\setcounter{table}{0}
\setcounter{equation}{0}
\renewcommand{\thepage}{S\arabic{page}}
\setcounter{page}{1}
\setcounter{section}{0}
\renewcommand{\thesection}{S\arabic{section}}
\ifdefined\theHsection\renewcommand{\theHsection}{S\arabic{section}}\fi
\begin{center}
\section*{Supplementary Materials for\\The quantum geometry of the chemical bond}
Gabrielle Carrel, Nishchhal Verma, Fatmag\"ul Katmer,\\
Leslie M. Schoop and Raquel Queiroz\\[4pt]
{\small Corresponding authors: lschoop@princeton.edu;\\
raquel.queiroz@columbia.edu}
\end{center}
\subsection*{This PDF includes}
Materials and Methods (sections~\ref{app:materials_benchmarks} to~\ref{app:tb_models});
Supplementary Text (sections~\ref{app:qgt_algebra} to~\ref{sec:chern-motifs-v15});
Figs.~\ref{fig:motifs-haldane-v15} and~\ref{fig:SI_benzene_fluctuation}; Tables~\ref{tab:relaxed_structures} to~\ref{tab:class_contributions}.

\clearpage
\subsection*{Materials and Methods}

\section{Electronic-structure calculations}
\label{app:materials_benchmarks}

Density functional theory calculations were performed with the Vienna Ab initio Simulation Package (VASP) 6.4.2, the PBE exchange--correlation functional, and the projector-augmented-wave method with the VASP-recommended PBE potentials (Na\_pv/Cl, C, Cs\_sv/I, Ge\_d/Te). All calculations were non-spin-polarized and did not include spin--orbit coupling. Ionic positions and lattice parameters were relaxed with a 550~eV plane-wave cutoff and Gaussian smearing of 0.02~eV until all forces were below $5\times10^{-3}$~eV/\AA; the relaxed cells are listed in Table~\ref{tab:relaxed_structures}. Static calculations on the relaxed structures used $\Gamma$-centered $k$-point meshes generated with VASPKIT\cite{gengEmpoweringMaterialsScience2025}: $15\times15\times15$ for NaCl and GeTe, $16\times16\times16$ for diamond, and $6\times9\times6$ for \CsI, with a plane-wave cutoff of 550~eV (400~eV for diamond), and electronic convergence criterion of $10^{-8}$~eV. 

Wannier functions were constructed with Wannier90 3.1.0\cite{pizziWannier90CommunityCode2020} on the same meshes, from atomic-orbital projections (Na $s,p$ and Cl $p$; C $s,p$; I $p$; Ge and Te $s,p$). The disentanglement windows are listed in Table~\ref{tab:settings}\cite{souzaMaximallyLocalizedEntangled2001}; the disentangled orbitals were used without further maximal localization (1000 disentanglement iterations and no localization iteration for NaCl, \CsI, and GeTe; one localization iteration for diamond). The resulting Wannier Hamiltonians and centers define the local orbitals $|\phi_\alpha\rangle$ of the main text. The \CsI\ model contains iodine $p$ orbitals only, so all \CsI\ quantities resolve the iodine manifold.

Crystal orbital bond indices were computed with LOBSTER 5.1.1\cite{maintzLOBSTERToolExtract2016} and the pbeVaspFit2015 basis from separate static calculations without symmetry reduction. To compare with the Wannier description, only the occupied states inside the Wannier energy window were included: for each atom pair the energy-resolved COBI was integrated from the lower edge of the Wannier outer window, $E_{\mathrm{start}}$, to the Fermi level,
\begin{equation}
  \mathrm{ICOBI}_{\mathrm{W}} = \int_{E_{\mathrm{start}}}^{E_F}\mathrm{COBI}(E)\,dE
  = \mathrm{ICOBI}(-\infty \to E_F) - \mathrm{ICOBI}(-\infty \to E_{\mathrm{start}}).
  \label{eq:icobi_window_si}
\end{equation}
For \CsI, Cs basis functions were included in the projection. For consistency with the iodine-only Wannier model, only I--I bond indices were summed. 

All tensors describe the Kohn--Sham determinant in the retained Wannier manifold and are given per retained electron, with the same spin convention in numerator and denominator. The bond indices of Table~\ref{tab:bondorder_compare} and bond fluctuations of Table \ref{tab:fig2-fluctuations} are evaluated in the primitive cell. Hoppings with $|t|\ge10^{-4}$~eV are included.  Table~\ref{tab:bondorder_compare} compares the shell-resolved bond-index partition of the metric trace, $\operatorname{Tr}g=\sum_{\rm pairs}b\,d^2/2N_e$ (section~\ref{app:qgt_algebra}), between the Wannier and LOBSTER descriptions.
\begin{table}[htbp]
\centering\small
\caption{\textbf{Relaxed structures.} Lattice parameters, angles $(\alpha,\beta,\gamma)$, and cell volumes of the primitive cells.}
\label{tab:relaxed_structures}
\begin{tabular}{lrrr}
\toprule
Material & $(a,b,c)$ (\AA) & angles (deg) & $V$ (\AA$^3$)\\
\midrule
NaCl & $(4.02653,4.02653,4.02653)$ & $(60,60,60)$ & 46.16113\\
Diamond & $(2.52636,2.52636,2.52636)$ & $(60,60,60)$ & 11.40171\\
\CsI & $(11.26904,7.24674,10.50050)$ & $(90,90,90)$ & 857.51179\\
GeTe & $(4.40278,4.40278,4.40278)$ & $(56.81262,56.81262,56.81262)$ & 55.90896\\
\bottomrule
\end{tabular}
\end{table}

\begin{table}[htbp]
\centering\small
\caption{\textbf{Wannier and LOBSTER settings.} Bands and Wannier functions (WF) per primitive cell, number of electrons $N_e$ in the Wannier manifold per primitive cell (both spins), DFT $k$-mesh, plane-wave cutoff, and Fermi level $E_F$ of the static calculation. The outer and frozen disentanglement windows used in the Wannier90 calculation are given. The windowed ICOBI of Eq.~\eqref{eq:icobi_window_si} is integrated from the lower edge of the outer window to $E_F$ of the LOBSTER calculation. The last column gives the absolute charge spilling of the LOBSTER projection.}
\label{tab:settings}\setlength{\tabcolsep}{5.5pt}
\begin{tabular}{lcccccccc}
\toprule
Material & Bands/WF & $N_e$ & DFT mesh & Cutoff (eV) & $E_F$ (eV) & Outer (eV) & Frozen (eV) & \shortstack{Charge\\Spilling (\%)} \\
\midrule
NaCl    & 36/7   & 6  & $15^3$             & 550 & $-0.70$ & $[-5, 18]$  & $[-5, 11]$  & 0.30 \\
Diamond & 12/8   & 8  & $16^3$             & 400 & $10.05$ & $[-19, 30]$ & $[-15, 25]$ & 0.98 \\
CsI$_3$ & 144/36 & 64 & $6\times9\times6$  & 550 & $0.68$  & $[-6, 3]$   & $[-5, 3]$   & 0.78 \\
GeTe    & 36/8   & 10 & $15^3$             & 550 & $5.99$  & $[-15, 15]$ & $[-10, 6]$  & 1.02 \\
\bottomrule
\end{tabular}
\end{table}

\begin{table}
    \centering\small
    \setlength{\tabcolsep}{3pt}
    \begin{tabular}{llccccccc}
    \hline
  & Atoms & $d$ (\AA) & Mult. & \multicolumn{2}{c}{Wannier} & \multicolumn{2}{c}{LOBSTER} & $b_\mathrm{W}$ / $b_\mathrm{L}$ \\
  & & & & $\sum b\,d^2$ (\AA$^2$) & \% & $\sum b\,d^2$ (\AA$^2$) & \% & \\
    \hline
  NaCl &&&&&&&& \\
  & Cl--Na & 2.85 & 6 & 3.72 & 85.2 & 2.42 & 65.3 & 0.0763 / 0.0499 \\
  & Cl--Na & 6.37 & 24 & 0.17 & 3.8 & 0.10 & 2.6 & 0.0002 / 0.0001 \\
  & Cl--Cl & 4.03 & 6 & 0.13 & 3.0 & 0.56 & 15.0 & 0.0014 / 0.0057 \\
  & Na--Na & 5.69 & 3 & 0.12 & 2.7 & 0.03 & 0.8 & 0.0012 / 0.0003 \\
  & Na--Na & 4.03 & 6 & 0.04 & 0.8 & 0.01 & 0.4 & 0.0004 / 0.0001 \\
  & Other shells & -- & -- & 0.20 & 4.5 & 0.59 & 15.9 & \\
  & Total ($d \le 20$~\AA) & & & 4.37 & 100.0 & 3.72 & 100.0 & \\
  & \multicolumn{8}{l}{$\mathrm{Tr}\,g$ at 20~\AA: Wannier 0.364~\AA$^2$, LOBSTER 0.310~\AA$^2$ ($N_e = 6$)} \\
  \hline
  Diamond &&&&&&&& \\
  & C--C & 1.55 & 4 & 9.05 & 73.6 & 9.10 & 74.6 & 0.9416 / 0.9504 \\
  & C--C & 3.89 & 12 & 1.65 & 13.4 & 1.72 & 14.1 & 0.0091 / 0.0095 \\
  & C--C & 5.05 & 12 & 0.27 & 2.2 & 0.16 & 1.3 & 0.0009 / 0.0005 \\
  & C--C & 2.96 & 12 & 0.26 & 2.1 & 0.20 & 1.6 & 0.0025 / 0.0019 \\
  & C--C & 6.38 & 24 & 0.24 & 2.0 & 0.22 & 1.8 & 0.0002 / 0.0002 \\
  & Other shells & -- & -- & 0.83 & 6.7 & 0.80 & 6.5 & \\
  & Total ($d \le 20$~\AA) & & & 12.30 & 100.0 & 12.20 & 100.0 & \\
  & \multicolumn{8}{l}{$\mathrm{Tr}\,g$ at 20~\AA: Wannier 0.769~\AA$^2$, LOBSTER 0.762~\AA$^2$ ($N_e = 8$)} \\
  \hline
  CsI$_3$ (I--I pairs) &&&&&&&& \\
  & I--I & 5.92 & 4 & 38.02 & 44.2 & 37.81 & 44.8 & 0.2712 / 0.2697 \\
  & I--I & 2.90 & 4 & 19.57 & 22.7 & 16.39 & 19.4 & 0.5819 / 0.4866 \\
  & I--I & 3.02 & 4 & 15.62 & 18.1 & 13.81 & 16.4 & 0.4281 / 0.3790 \\
  & I--I & 6.94 & 4 & 3.66 & 4.3 & 3.20 & 3.8 & 0.0190 / 0.0166 \\
  & I--I & 9.95 & 4 & 3.54 & 4.1 & 3.67 & 4.4 & 0.0089 / 0.0093 \\
  & Other shells & -- & -- & 5.64 & 6.6 & 9.56 & 11.3 & \\
  & Total ($d \le 20$~\AA) & & & 86.06 & 100.0 & 84.44 & 100.0 & \\
  & \multicolumn{8}{l}{$\mathrm{Tr}\,g$ at 20~\AA: Wannier 0.673~\AA$^2$, LOBSTER 0.660~\AA$^2$ ($N_e = 64$)} \\
  \hline
  GeTe &&&&&&&& \\
  & Ge--Te & 2.84 & 3 & 14.91 & 27.6 & 15.35 & 26.8 & 0.6161 / 0.6337 \\
  & Ge--Te & 3.26 & 3 & 5.35 & 9.9 & 5.71 & 10.0 & 0.1678 / 0.1791 \\
  & Ge--Te & 8.91 & 6 & 4.14 & 7.7 & 3.42 & 6.0 & 0.0087 / 0.0072 \\
  & Ge--Te & 6.57 & 6 & 2.22 & 4.1 & 2.17 & 3.8 & 0.0086 / 0.0084 \\
  & Ge--Ge & 6.08 & 3 & 2.17 & 4.0 & 3.30 & 5.8 & 0.0195 / 0.0298 \\
  & Other shells & -- & -- & 25.16 & 46.6 & 27.32 & 47.7 & \\
  & Total ($d \le 20$~\AA) & & & 53.93 & 100.0 & 57.27 & 100.0 & \\
  & \multicolumn{8}{l}{$\mathrm{Tr}\,g$ at 20~\AA: Wannier 2.697~\AA$^2$, LOBSTER 2.864~\AA$^2$ ($N_e = 10$)} \\
  \hline
    \end{tabular}
    \caption{\textbf{Shell-resolved bond-index partition of the quantum metric}, $\mathrm{Tr}\,g = \sum_\mathrm{pairs} b\,d^2 / 2N_e$, from the Wannier models and from LOBSTER.  The trace is truncated at 20~\AA\ for all systems. Measured against the full traces of Table~\ref{tab:fig2-fluctuations}, the Wannier 20~\AA\ sums recover 99.0, 100.0, 99.4 and 86.8\% for NaCl, diamond, \CsI\ and GeTe. Mult.\ is the number of pairs of the shell per primitive cell. The five shells with the largest Wannier contribution are explicitly listed. The last column gives the bond index of the pair from the Wannier and LOBSTER calculations}. 
\label{tab:bondorder_compare}
\end{table}

\begin{table}[htbp]
\centering
\caption{\textbf{Individual fluctuations shown in Fig.~\ref{fig:bonding_combined}E--H.} Sites are listed in the order
$1$--$2$--$3$--$4$ of the fluctuation pattern (Eq.~\ref{eq:bond_resolved_velocity_tensor_main} and Fig.~\ref{fig:bonding_combined}\textbf{A--D}); primes mark distinct atoms
of the same element. In CsI$_3$, I$_\mathrm{c}$ is the central iodine of the I$_3^-$ unit and I$_1$,
I$_3$ are the terminal iodines at 3.02 and 2.90~\AA; in GeTe, the short and long Ge--Te bonds are
2.84 and 3.26~\AA. The multiplicity is the number of symmetry-equivalent terms per primitive cell,
with a loop and its reverse traversal counted separately; the contribution is their sum. All values
are in \AA$^2$ per retained electron; $\mathrm{Tr}\,g$ is the total metric trace of each material shown in Fig.~\ref{fig:bonding_combined}\textbf{I}.}
\label{tab:fig2-fluctuations}
\begin{tabular}{lcllcr}
\toprule
Material & $\mathrm{Tr}\,g$ (\AA$^2$) & Fluctuation & Sites & Multiplicity & Contribution (\AA$^2$) \\
\midrule
NaCl    & 0.368 & $\alpha\beta\beta\alpha$  & Cl--Na--Na--Cl & 6 & 0.336 \\
\midrule
Diamond & 0.769 & $\alpha\beta\beta\alpha$  & C--C$'$--C$'$--C & 8 & 0.405 \\
        &       & $\alpha\beta\alpha\beta$  & C--C$'$--C--C$'$ & 8 & 0.298 \\
\midrule
CsI$_3$ & 0.677 & $\alpha\beta\beta\gamma$  & I$_1$--I$_\mathrm{c}$--I$_\mathrm{c}$--I$_3$ & 8 & 0.212 \\
        &       & $\alpha\beta\beta\alpha$  & I$_1$--I$_\mathrm{c}$--I$_\mathrm{c}$--I$_1$ & 4 & 0.143 \\
        &       & $\alpha\beta\beta\alpha$  & I$_3$--I$_\mathrm{c}$--I$_\mathrm{c}$--I$_3$ & 4 & 0.134 \\
        &       & $\alpha\beta\alpha\beta$  & I$_\mathrm{c}$--I$_3$--I$_\mathrm{c}$--I$_3$ & 8 & 0.047 \\
        &       & $\alpha\beta\gamma\alpha$ & I$_\mathrm{c}$--I$_1$--I$_3$--I$_\mathrm{c}$ & 8 & $-0.041$ \\
\midrule
GeTe    & 3.105 & $\alpha\beta\beta\alpha$  & Te--Ge--Ge--Te (short bond) & 3 & 0.796 \\
        &       & $\alpha\beta\beta\alpha$  & Te--Ge--Ge--Te (long bond)  & 3 & 0.445 \\
        &       & $\alpha\beta\beta\gamma$  & Te--Ge--Ge--Te$'$           & 6 & 0.372 \\
        &       & $\alpha\beta\gamma\alpha$ & Te--Ge--Ge$'$--Te           & 6 & 0.332 \\
        &       & $\alpha\beta\gamma\delta$ & Te--Ge--Ge$'$--Te$'$        & 6 & 0.210 \\
\bottomrule
\end{tabular}
\end{table}

\clearpage

\section{Numerical evaluation of the bond-resolved tensor (Figure \ref{fig:bonding_combined})}
\label{app:combined_figure_v15}

The velocity-resolved tensor of the main text (Eq.~\ref{eq:bond_resolved_velocity_tensor_main}) is evaluated from the Wannier Hamiltonian $H(\mathbf k)$, written in the physical embedding so that $\nabla_{k_\mu}H=\hbar v^\mu$ gives the correct velocity operator. The Hamiltonian is diagonalized with matrix $U$ whose elements are $U_{\alpha n}(\mathbf k)=\langle\phi_\alpha|\psi_n\rangle$.  The position is taken to be diagonal in the local orbital basis, so a hopping $t_{\alpha\beta}(\mathbf R)$ to the cell at $\mathbf R$ contributes the velocity matrix element \begin{equation}
v_{\alpha\beta}^\mu(\mathbf R) = -\frac{i}{\hbar}R^\mu_{\alpha\beta}\, t_{\alpha\beta}, \qquad R_{\alpha\beta}^\mu = r_\alpha^\mu-r_\beta^\mu
\label{eq:velocity_matrix_element}
\end{equation} 
where $\mathbf r_\alpha$ is the wannier center of orbital $\alpha$. 
All four models are gapped at the $10^{-4}$~eV hopping cutoff, so the lowest $N_{\rm occ}$ bands are
occupied at every $\mathbf k$. The quantum metric then reduces to the zero-temperature sum over
occupied $n$ and empty $m$ of section~\ref{app:decomposition_equivalence}, evaluated on
half-step-shifted $n_k^3$ meshes with $n_k = 48$, 46, 15, and 42 for NaCl, diamond, \CsI, and GeTe,
respectively, and normalized by the occupied counts, $N_{\rm occ}=N_e/2$ occupied bands per primitive cell with $N_e$ from Table~\ref{tab:settings}.

Terms are classified by physical site as in section~\ref{app:class_physics}, so translated copies of an atom
are distinct centers. The energy denominator couples $n$ and $m$. Writing it as a time integral,
$(\epsilon_m-\epsilon_n)^{-2}=\int_0^\infty t\,e^{-(\epsilon_m-\epsilon_n)t}\,dt$, allows Eq.~\eqref{eq:app_four_index} to be split 
into a filled-state factor $\rho(t)=\sum_{n\in{\rm occ.}}e^{(\epsilon_n-\mu_0)t}P_n$ and an
empty-state factor $\bar\rho(t)=\sum_{m\in{\rm unocc.}}e^{-(\epsilon_m-\mu_0)t}P_m$, with $\mu_0$ at
mid-gap, as in Eq.~\eqref{eq:Qt}
\begin{equation}
  Q^{\mu\nu}_{\alpha\beta\gamma\delta}=-\frac{1}{N_e}\,H_{\alpha\beta}R^\mu_{\alpha\beta}\,
  H_{\gamma\delta}R^\nu_{\gamma\delta}\int_0^\infty t\,\rho_{\delta\alpha}(t)\,\bar\rho_{\beta\gamma}(t)\,dt 
\label{eq:Qt}
\end{equation}
These filled and empty state factors can be written as  Fourier transforms of the Bloch eigenvectors $U_{\alpha n}(\mathbf k)$, allowing real-space resolution of the orbital-resolved quantum metric. 
\begin{equation}
\begin{aligned}
  \rho_{\alpha\beta}(t)&=\frac{1}{N_k}\sum_{\mathbf k}e^{i\mathbf k\cdot(\mathbf r_\alpha-\mathbf r_\beta)}
  \sum_{n\in{\rm occ.}}e^{(\epsilon_{n\mathbf k}-\mu_0)t}\,U_{\alpha n}(\mathbf k)U^*_{\beta n}(\mathbf k),\\
  \bar\rho_{\alpha\beta}(t)&=\frac{1}{N_k}\sum_{\mathbf k}e^{i\mathbf k\cdot(\mathbf r_\alpha-\mathbf r_\beta)}
  \sum_{m\in{\rm unocc.}}e^{-(\epsilon_{m\mathbf k}-\mu_0)t}\,U_{\alpha m}(\mathbf k)U^*_{\beta m}(\mathbf k),
\end{aligned}
  \label{eq:rhot}
\end{equation}

All patterns with three or fewer distinct centers are summed in this form to create the fluctuation class decomposition of Fig.~\ref{fig:bonding_combined}\textbf{I}, while the
four-center class is the remainder of the full tensor. Individual four-center fluctuations , like the GeTe term in ~~Table \ref{tab:fig2-fluctuations} are enumerated directly in real space. The fluctuation classes are listed in Table~\ref{tab:class_contributions}. Table~\ref{tab:fig2-fluctuations} lists the individual contributions of symmetry equivalent fluctuations plotted in Fig.~\ref{fig:bonding_combined}\textbf{E--H}.
\begin{table}[htbp]
\centering\small
\caption{\textbf{Contributions of the site patterns to the metric trace}, in \AA$^2$ per retained electron, in Fig.~\ref{fig:bonding_combined}\textbf{I}. The eight patterns that vanish in the orbital-center approximation (Eq.~\eqref{eq:physical_site_classes_si}) are omitted. Signed contributions can exceed the total because other patterns cancel them.}
\label{tab:class_contributions}
\begin{tabular}{llrrrr}
\toprule
Class & Pattern & NaCl & Diamond & \CsI & GeTe\\
\midrule
Two-center polarization & $\alpha\beta\beta\alpha$ & 0.406 & 0.503 & 0.357 & 1.648\\
Two-center transfer & $\alpha\beta\alpha\beta$ & 0.011 & 0.298 & 0.068 & 0.188\\
Three-center & $\alpha\beta\beta\gamma$ & $-$0.046 & $-$0.051 & 0.233 & 0.075\\
 & $\alpha\beta\gamma\alpha$ & 0.018 & 0.052 & 0.016 & 0.554\\
 & $\alpha\beta\alpha\gamma$ & $-$0.012 & $-$0.001 & $-$0.002 & $-$0.024\\
 & $\alpha\beta\gamma\beta$ & $-$0.012 & $-$0.001 & $-$0.002 & $-$0.024\\
Four-center & $\alpha\beta\gamma\delta$ & 0.002 & $-$0.030 & 0.006 & 0.687\\
\bottomrule
\end{tabular}
\end{table}

\section{Fluctuation spectrum and spatial range (Figure \ref{fig:linked_channel_modes_v15})}
\label{app:physical_image_channel_modes_v15}
\label{app:svd_modes}

\paragraph{Fluctuation spectrum.}
As in the main text, we read the four-index tensor of Eq.~\ref{eq:bond_resolved_velocity_tensor_main} as a matrix between pairs of orbitals.
Summing over the Cartesian components gives the bond--bond fluctuation matrix
\begin{equation}
K_{(\alpha\beta),(\gamma\delta)}=\sum_{\mu\in\mathcal I}Q^{\mu\mu}_{\alpha\beta\delta\gamma}
=\sum_\ell\mathsf B_{(\alpha\beta),\ell}\,\mathsf B_{(\gamma\delta),\ell}^*,
\label{eq:image_mode_metric_v15}
\end{equation}
where $\mathsf B_{(\alpha\beta),\ell}=\hbar\langle\psi_n|v^\mu_{\alpha\beta}|\psi_m\rangle/[(\epsilon_m-\epsilon_n)\sqrt{N_e}]$ resolves the transition-dipole matrix elements of the main text into directed-bond contributions, normalized by $\sqrt{N_e}$. Summing over bonds gives $\sum_{\alpha\beta}\mathsf B_{(\alpha\beta),\ell}=\langle\psi_n|\mathcal D^\mu|\psi_m\rangle/\sqrt{N_e}$, explicitly relating these amplitudes to the transition-dipole operator of the main text. The velocity matrix elements are given in Eq.~\eqref{eq:velocity_matrix_element}. The composite index $\ell=(\mathbf k,\mu,n,m)$ collects the momentum $\mathbf k$, Cartesian component $\mu$, occupied band $n$, and empty band $m$. The sum over $\ell$ includes all these indices, so $K$ is a momentum-integrated matrix.
Its eigendecomposition
\begin{equation}
K=\sum_s\lambda_su_su_s^\dagger
\end{equation}
gives the normal modes $u_s$ of the bond fluctuations and the fluctuation spectrum $\lambda_s$, which is nonnegative because $K=\mathsf B\mathsf B^\dagger$ is positive semidefinite. Figure~\ref{fig:linked_channel_modes_v15}\textbf{A} shows the normalized eigenvalues $\lambda_s/\sum_s\lambda_s$ for the four materials.

\paragraph{Spatial range.}
Since $K$ is constructed using the orbital pairs of the chosen unit cell, the normal modes are not suited to define a macroscopic spatial range.
A mode typically contains symmetry equivalent copies of a bond making it difficult to identify the minimal repeating unit or motif. 
We obtain the range instead from a trial fluctuation placed on one bonding unit, which we project back into real space through the momentum-resolved amplitudes.

We choose the bonding unit from chemical intuition. For NaCl it is one Na--Cl pair, for diamond the four C--C bonds around one carbon, for \CsI\ the three I--I pairs of one I$_3^-$ unit, and for GeTe the six Ge--Te bonds around one germanium. In the $2\times2\times2$ supercells, and in the unit cell of \CsI, these bonds are far from their periodic copies.

We diagonalize the block of $K$ restricted to the orbital pairs $S$ of the bonding unit, $K_S v_j = h_j v_j$. 
The vector $x_j=Kv_j/\sqrt{h_j}$, with $v_j$ set to zero outside $S$, contains the correlations of the eigenvector $v_j$ with all orbital pairs in the cell, and $|\boldsymbol1^\dagger x_j|^2$ is its contribution to the metric trace. 
Here $\boldsymbol1$ is the vector with all entries equal to one, so that $\boldsymbol1^\dagger$ sums over all orbital pairs and $\boldsymbol1^\dagger K\boldsymbol1=\sum_\mu g^{\mu\mu}$.
We take the $v_j$ with the largest contribution as trial. This is not always the one with the largest $h_j$, because amplitudes of opposite sign can cancel in the metric sum.

We project the momentum-resolved amplitudes onto the trial, $w_j=v_j^\dagger\mathsf B/\sqrt{h_j}$, which gives the normalized amplitude $w_j(\mathbf k,\mu,n,m)$ of each transition $n\to m$ at momentum $\mathbf k$. In the orbital basis these amplitudes become
\begin{equation}
\Psi_{ab}^\mu(\mathbf k)=\sum_{n,m}U_{an}(\mathbf k)\,w_j(\mathbf k,\mu,n,m)\,U_{bm}^*(\mathbf k),
\label{eq:right_mode_image_kernel_v15}
\end{equation}
with $U_{an}=\langle\phi_a|\psi_n\rangle$. Orbital $a$ carries the occupied side of each transition and orbital $b$ the empty side, and $\Psi$ does not depend on the band phases. We then Fourier transform $\Psi_{ab}^\mu(\mathbf k)$ to $\Psi_{ab}^\mu(\mathbf R)$.
 
We sum $|\Psi_{ab}^\mu(\mathbf R)|^2$ over the orbitals of each atom and over $\mu$ to obtain the weight $W_{AB}(\mathbf R)$ of transitions from atom $A$ in the home cell to atom $B$ in the cell at $\mathbf R$. 
With $d_{AB}(\mathbf R)$ the distance between the two atoms, $r_{90}$ is the smallest distance that contains $90\%$ of the weight,
\begin{equation}
F(r)=\sum_{d_{AB}(\mathbf R)\le r}W_{AB}(\mathbf R),\qquad r_{90}=\min\{r:F(r)\ge0.9\}.
\end{equation}
Since $F$ only grows at interatomic distances, $r_{90}$ is always a neighbor-shell distance. Figure~\ref{fig:linked_channel_modes_v15}\textbf{B} shows $F(r)$ for the four materials.

\section{Tight-binding models (Figure \ref{fig:loop_chemistry_summary})}
\label{app:tb_models}
\label{app:tb_loop_response}

For the molecular graphs, open flakes, and the periodic Haldane lattice of Figure~\ref{fig:loop_chemistry_summary}, the occupied projector and the transition spectrum are computed from the same Hamiltonian. For spinless honeycomb graphs with one orbital per site,
\begin{equation}
H=-t\sum_{\langle ij\rangle}(c_i^\dagger c_j+\mathrm{h.c.})
-t'\sum_{\langle\!\langle ij\rangle\!\rangle}
(e^{-i\phi\nu_{ij}}c_i^\dagger c_j+\mathrm{h.c.}),
\label{eq:haldane_main}
\end{equation}
counting each unoriented bond once, with a fixed orientation in the second sum; $\nu_{ij}=\pm1$ records the circulation of the two-step path through the shared nearest neighbor, and $\nu_{ji}=-\nu_{ij}$. The nearest-neighbor distance is $a$ and the cell area $A_c=3\sqrt3 a^2/2$. All geometries use $t'=t$, $\phi=\pi/2$, equal sublattice energies, and half filling, $N_e=N/2$ for a finite graph and $N_{e,c}=1$ for the periodic cell. The complex second-neighbor hopping imposes a staggered flux with zero net flux per cell; it breaks time reversal and opens the topological gap of the periodic lattice. The zigzag ribbon of Fig.~\ref{fig:loop_chemistry_summary}\textbf{I} is periodic along the edge with period $\sqrt3\,a$ and contains 100 zigzag chains across its width, 200 sites per repeating cell.

The spectra of Fig.~\ref{fig:loop_chemistry_summary}\textbf{D--F} are $-2\operatorname{Im}\mathcal Q^{xy}(E)$ per electron from Eq.~\eqref{eq:app_spectral_density}, in units of $a^2/t$. Each transition is broadened by a Gaussian of standard deviation $\eta=0.075\,t$ on an energy grid of spacing $10^{-3}\,t$, mirrored at $E=0$ so that the weight at $E\ge0$ is preserved; the same broadening is used for all geometries of Figs.~\ref{fig:loop_chemistry_summary}\textbf{D--F}. The molecular graphs and flakes are diagonalized exactly, with every occupied-to-empty pair retained, and the periodic bulk is summed on a $\Gamma$-centred $480\times480$ $k$-mesh. The ribbon bands of Fig.~\ref{fig:loop_chemistry_summary}\textbf{I} are computed at 301 points across the edge Brillouin zone. Integrating a longitudinal spectrum gives the corresponding metric component; integrating the transverse spectrum gives the curvature per electron, which cancels over the complete spectrum of a finite graph (section~\ref{app:molecular}) while the periodic bulk retains a nonzero integral. Weighting by excitation energy gives the transverse absorptive conductivity through Eq.~\eqref{eq:app_sigma_channel}; finite graphs have an intrinsic polarizability, and a conductivity requires a specified area.

\label{app:ring_level_inversion}
For the six-site ring, the orbitals carry rotation labels $m=0,\ldots,5$ modulo six. Complex hopping splits the pairs $(1,5)$ and $(2,4)$, and circular dipole transitions obey $\Delta m=\pm1$ modulo six. For the half-filled ring with the directed second-neighbor term $-t'e^{i\phi}c_{j+2}^\dagger c_j+\mathrm{h.c.}$, the positive peaks of Fig.~\ref{fig:loop_chemistry_summary}\textbf{G,H} are the transitions $4\to5$ and $1\to2$ ($\Delta m=+1$) and the negative peak contains the coincident transitions $0\to5$ and $4\to3$ ($\Delta m=-1$); their signed areas cancel. Importantly, the appearance of balancing positive and negative contributions appear from the inversion of $m=4$ and $m=5$ from the occupied to the unoccupied sectors respectively. This level inversion is the percursor of the band inversion in the periodic net leading to the total Chern number and topological edge modes.  At $t'=0$ the $m=\pm1$ and $m=\pm2$ pairs are the degenerate $e_{1g}$ and $e_{2u}$ orbitals, and time-reversal symmetry gives equal-energy, equal-strength transitions of opposite circular polarization, so their contributions cancel at every energy.

\clearpage
\subsection*{Supplementary Text}

\section{Bond indices and the quantum geometric tensor}
\label{app:qgt_algebra}

Throughout the Supplementary Text we consider a gapped ground state of non-interacting electrons at zero temperature, so that the one-particle density matrix is equal to the projector onto the occupied states, $P=\sum_{n\in\mathrm{occ.}}|\psi_n\rangle\langle\psi_n|$. 
A correlated many-body ground state can contain additional correlations that make the density matrix different than the single-particle projector.

The local orbitals $|\phi_\alpha\rangle$ form an orthonormal set that defines the projectors $\Pi_\alpha=|\phi_\alpha\rangle\langle\phi_\alpha|$. 
In this local-orbital basis the matrix elements of $P$ and $Q$ are the $\rho_{\alpha\beta}$ and $\bar\rho_{\alpha\beta}=\delta_{\alpha\beta}-\rho_{\alpha\beta}$, as mentioned in the main text.
In a crystal, $\alpha=(\mathbf T_\alpha,A_\alpha,a_\alpha)$ is typically a composite label for the unit cell $T_\alpha$, the atom within the cell $A_\alpha$, and the internal label $a_\alpha$ for orbital or spin.

The two- and three-center bond indices are
\begin{align}
b_{\alpha\beta}&=\operatorname{Tr}(\Pi_\alpha P\Pi_\beta P)
=|\rho_{\alpha\beta}|^2,\nonumber\\
b_{\alpha\beta\gamma}&=\operatorname{Tr}(\Pi_\alpha P\Pi_\beta P\Pi_\gamma P)
=\rho_{\alpha\beta}\rho_{\beta\gamma}\rho_{\gamma\alpha},
\qquad \sum_\gamma b_{\alpha\beta\gamma}=b_{\alpha\beta},
\label{eq:three-center_BI}
\end{align}
where the sum rule follows from $P^2=P$. The three-center index is invariant under cyclic permutations of $\alpha\to\beta\to\gamma$ by cyclicity of the trace.

The transition dipole operator is (with the electron charge set to one)
\begin{equation}
\mathcal D^\mu=i[\hat r^\mu,P]=i(Q\hat r^\mu P-P\hat r^\mu Q), \qquad Q = 1-P,
\end{equation}
so that $P\mathcal D^\mu P=Q\mathcal D^\mu Q=0$ and $\langle\psi_n|\mathcal D^\mu|\psi_m\rangle=-i\langle\psi_n|\hat r^\mu|\psi_m\rangle$ for occupied $n$ and empty $m$. The quantum geometric tensor can now be written as
\begin{equation}
\mathcal Q^{\mu\nu}=\frac{1}{N_e}\operatorname{Tr}(P\mathcal D^\mu\mathcal D^\nu)
=\frac{1}{N_e}\operatorname{Tr}(P\hat r^\mu Q\hat r^\nu P),
\label{eq:sm_qgt}
\end{equation}
where $N_e$ is the number of electrons.

To express the tensor through bond indices, we use the orbital-center approximation where we write $\hat r^\mu\simeq\sum_\alpha r_\alpha^\mu\Pi_\alpha$ with $R^\mu_{\alpha\beta}=r_\alpha^\mu-r_\beta^\mu$. 
This neglects the finite extent of each orbital and the on-site dipole matrix elements between orbitals of the same atom. We thereby get the QGT written in terms of bond indices
\begin{equation}
\mathcal D^\mu_{\alpha\beta}=iR^\mu_{\alpha\beta}\rho_{\alpha\beta},\qquad
\mathcal Q^{\mu\nu}=-\frac{1}{N_e}\sum_{\alpha\beta\gamma}
R^\mu_{\beta\gamma}R^\nu_{\gamma\alpha}b_{\alpha\beta\gamma}.
\label{eq:qgt_loop_derivation}
\end{equation}

For the symmetric part, we can use $\operatorname{Tr}(\mathcal D^\mu\mathcal D^\nu)=2\operatorname{Re}\operatorname{Tr}(P\mathcal D^\mu\mathcal D^\nu)$ to get
\begin{equation}
g^{\mu\nu}=\operatorname{Re}\mathcal Q^{\mu\nu}
=\frac{1}{2N_e}\sum_{\alpha\beta}b_{\alpha\beta}
R^\mu_{\alpha\beta}R^\nu_{\alpha\beta},
\label{eq:qgt_two_center_sym}
\end{equation}
so that the quantum metric, $g^{\mu\nu}$, is a sum over directed pairs. The antisymmetric part retains the loop phase,
\begin{align}
\mathcal A^{\mu\nu}_{\alpha\beta\gamma}
&=\tfrac12(R^\mu_{\beta\gamma}R^\nu_{\gamma\alpha}
-R^\nu_{\beta\gamma}R^\mu_{\gamma\alpha}),\nonumber\\
\Omega^{\mu\nu}=-2\operatorname{Im}\mathcal Q^{\mu\nu}
&=\frac{4}{N_e}\sum_\gamma\sum_{\alpha<\beta}
\mathcal A^{\mu\nu}_{\alpha\beta\gamma}\operatorname{Im}b_{\alpha\beta\gamma},
\label{eq:Omega_three_center}
\end{align}
where $\mathcal A^{\mu\nu}_{\alpha\beta\gamma}$ tracks the oriented area of the triangle formed by the orbital centers of $\alpha$, $\beta$ and $\gamma$, projected onto the $\mu\nu$ plane.

\section{Electron counting and periodic conventions}
\label{app:normalization}

For a molecule, Eq.~\eqref{eq:sm_qgt} is a full trace. For an infinite crystal it denotes the trace per cell divided by $N_{e,c}$, the retained occupied electron count per cell. In terms of Bloch states with $P_{\mathbf k} = \sum_{ (n, {\mathbf k}) \in {\rm occ.} } |\psi_{n, {\mathbf k} } \rangle \langle \psi_{n, {\mathbf k} }|$, we have
\begin{equation}
q^{\mu\nu}(\mathbf k)=\operatorname{tr}
[P_{\mathbf k}(\partial_{k_\mu}P_{\mathbf k})(\partial_{k_\nu}P_{\mathbf k})],
\qquad
\mathcal Q^{\mu\nu}=\frac{1}{N_{e,c}}
\int_{\mathrm{BZ}}\frac{d^dk}{V_{\mathrm{BZ}}}\,q^{\mu\nu}(\mathbf k).
\label{eq:periodic_qgt_si}
\end{equation}
We note that the per-electron normalization considered here differs from the Wiberg index by factors of spin degeneracy.

Both $g$ and $\Omega$ have units of length squared per electron. In two dimensions,
\begin{equation}
C=\frac{1}{2\pi}\int_{\mathrm{BZ}}d^2k\,[-2\operatorname{Im}q^{xy}]
=2\pi n_e\Omega^{xy},\qquad n_e=\frac{N_{e,c}}{A_c},
\label{eq:cell_chern_normalization_v15}
\end{equation}
with $A_c$ the primitive-cell area and $V_{\mathrm{BZ}}=(2\pi)^2/A_c$, so the Brillouin-zone average contributes the factor $2\pi N_{e,c}/A_c$ that converts the per-electron curvature into the dimensionless invariant.

\section{The quantum geometric tensor of a finite system is real}
\label{app:molecular}
\label{app:bounded_position}

For a finite system the position operator is bounded and its components commute $[\hat r^\mu, \hat r^\nu] = 0$. Using $Q=1-P$, we can write the QGT as
\begin{equation}
\mathcal Q^{\mu\nu}
=\frac{1}{N_e}\operatorname{Tr}(P\hat r^\mu\hat r^\nu)
-\frac{1}{N_e}\operatorname{Tr}(P\hat r^\mu P\hat r^\nu).
\label{eq:bounded_qgt_trace_si}
\end{equation}
to see that the first term is symmetric in $\mu\nu$ because the position components commute. Cyclicity of the trace makes the second term symmetric as well. Hermiticity then gives
\begin{equation}
\operatorname{Im}\mathcal Q^{\mu\nu}=0
\label{eq:bounded_qgt_real_si}
\end{equation}
which holds for all finite systems. 

Note that individual triangles and site-restricted sums can remain nonzero as the arguments above apply only to the complete sum over the system. 
Therefore, in a finite flake of a topological lattice the boundary is allowed to compensate for the bulk Chern marker (section~\ref{sec:chern-motifs-v15}). The argument assumes bounded position components. Projection onto an incomplete orbital basis violates the commutativity \cite{biancoMappingTopologicalOrder2011}.

\section{Transition and bond-index decompositions}
\label{app:decomposition_equivalence}

Let $H|\psi_n\rangle=\epsilon_n|\psi_n\rangle$ with $n\in\mathrm{occ.}$ and $m\in\mathrm{unocc.}$. From $v^\mu=(i/\hbar)[H,\hat r^\mu]$,
\begin{equation}
\langle\psi_n|v^\mu|\psi_m\rangle
=-\frac{i(\epsilon_m-\epsilon_n)}{\hbar}
\langle\psi_n|\hat r^\mu|\psi_m\rangle,\qquad
\langle\psi_n|\mathcal D^\mu|\psi_m\rangle
=\frac{\hbar\langle\psi_n|v^\mu|\psi_m\rangle}{\epsilon_m-\epsilon_n},
\label{eq:app_velocity_form}
\end{equation}
which is the velocity form of the transition dipole. With the directed orbital blocks $v^\mu_{\alpha\beta}=\Pi_\alpha v^\mu\Pi_\beta$, the bond-resolved tensor of the main text is
\begin{equation}
Q^{\mu\nu}_{\alpha\beta\gamma\delta}
=\frac{\hbar^2}{N_e}\sum_{\substack{n\in\mathrm{occ.}\\m\in\mathrm{unocc.}}}
\frac{\langle\psi_n|v^\mu_{\alpha\beta}|\psi_m\rangle
\langle\psi_m|v^\nu_{\gamma\delta}|\psi_n\rangle}
{(\epsilon_m-\epsilon_n)^2}.
\label{eq:velocity_partition_si}
\end{equation}
Since $\sum_{\alpha\beta}v^\mu_{\alpha\beta}=v^\mu$, its complete sum is
\begin{equation}
\begin{aligned}
\sum_{\alpha\beta\gamma\delta}Q^{\mu\nu}_{\alpha\beta\gamma\delta}
&=\frac{1}{N_e}\sum_{\substack{n\in\mathrm{occ.}\\m\in\mathrm{unocc.}}}
\langle\psi_n|\mathcal D^\mu|\psi_m\rangle
\langle\psi_m|\mathcal D^\nu|\psi_n\rangle\\
&=\frac{1}{N_e}\operatorname{Tr}(P\mathcal D^\mu Q\mathcal D^\nu)
=\mathcal Q^{\mu\nu},
\end{aligned}
\label{eq:bond_sum_qgt_main}
\end{equation}
using $Q\mathcal D^\nu P=\mathcal D^\nu P$. Expanding the trace in local orbitals gives Eq.~\eqref{eq:qgt_loop_derivation}, so the bond-index and the transition decompositions recover the same complete tensor, including both its real symmetric part (the metric) and its imaginary antisymmetric part (the curvature).

The two decompositions do not agree term by term. For the local dipole blocks $\Pi_\alpha\mathcal D^\mu\Pi_\beta$, with matrix elements $\mathcal D^\mu_{\alpha\beta}=iR^\mu_{\alpha\beta}\rho_{\alpha\beta}$ as in Eq.~\eqref{eq:qgt_loop_derivation},
\begin{equation}
\frac{1}{N_e}\operatorname{Tr}(P\,\Pi_\alpha\mathcal D^\mu\Pi_\beta\,\Pi_\gamma\mathcal D^\nu\Pi_\delta)
=\frac{\rho_{\delta\alpha}}{N_e}
 \mathcal D^\mu_{\alpha\beta}\mathcal D^\nu_{\gamma\delta}\delta_{\beta\gamma},
\end{equation}
so only terms with coinciding middle labels survive. 
The velocity partition instead inserts 
\begin{equation}
\sum_{m\in\mathrm{unocc.}} \dfrac{|\psi_m\rangle\langle\psi_m|}{(\epsilon_m-\epsilon_n)^2}
\end{equation}
between the two blocks; this energy-weighted operator is not an identity operator and its $\beta\gamma$ element can be nonzero for distinct orbitals, so the middle labels need not coincide even after integrating over all excitation energies. The two decompositions assign different shares to a given bond while summing to the same total, in a similar way that Mulliken and Hirshfeld population analyses assign different numbers of electrons to an atom while summing to the same total.

\subsection{Orbital-index patterns}
\label{app:class_physics}

In the orbital-center approximation, $v^\mu_{\alpha\beta}=-(i/\hbar)H_{\alpha\beta}R^\mu_{\alpha\beta}|\phi_\alpha\rangle\langle\phi_\beta|$, the resonance integral times the bond vector, and
\begin{equation}
Q^{\mu\nu}_{\alpha\beta\gamma\delta}
=-\frac{1}{N_e}\sum_{\substack{n\in\mathrm{occ.}\\m\in\mathrm{unocc.}}}
\frac{(P_n)_{\delta\alpha}(P_m)_{\beta\gamma}
 H_{\alpha\beta}R^\mu_{\alpha\beta}H_{\gamma\delta}R^\nu_{\gamma\delta}}
{(\epsilon_m-\epsilon_n)^2},\qquad
P_n=|\psi_n\rangle\langle\psi_n|,
\label{eq:app_four_index}
\end{equation}
Each term correlates two transition amplitudes of the same energy occupied-to-empty transition, one of them complex conjugated: $(P_n)_{\delta\alpha}$ carries the occupied-state weight on the two donor orbitals $\alpha$ and $\delta$, and $(P_m)_{\beta\gamma}$ the empty-state weight on the two acceptor orbitals $\beta$ and $\gamma$. Summed over all occupied $n$ and empty $m$, these factors become the density-matrix elements $\rho_{\delta\alpha}$ and $\bar\rho_{\beta\gamma}$ that label the connecting lines of Figure~\ref{fig:bond_ladder}. 

In the orbital-center approximation, velocity blocks with $R_{\alpha\beta}=0$ vanish. The surviving patterns are\begin{equation}\underbrace{\alpha\beta\alpha\beta,\ \alpha\beta\beta\alpha}_{\text{two-center}},\qquad \underbrace{\alpha\beta\alpha\gamma,\ \alpha\beta\beta\gamma,\ \alpha\beta\gamma\alpha,\ \alpha\beta\gamma\beta}_{\text{three-center}},\qquad \underbrace{\alpha\beta\gamma\delta}_{\text{four-center}}.\label{eq:physical_site_classes_si}\end{equation}Table~\ref{tab:class_contributions} gives their contributions.

\section{Hypervalent and metavalent bonding}
\label{app:hypervalent_metavalent}
Both analyses in Section ~\ref{sec:time_optical_response} show that the character of electronic transfers can be connected to bonding types and to how far bonds extend. They also help us understand the difference, or lack thereof, between hypervalent and metavalent bonding. Hypervalent bonding is most commonly described in molecules, and our example shows how the delocalization is retained in the molecular unit. Hoffmann and coworkers\cite{a.papoianHypervalentBondingOne2000a} extended the discussion to covalent nets; here such bonds can surely also have longer range. However, so far hypervalent bonding in solids has only been described between elements of the same type, such that there is no electronegativity difference. As a consequence, hypervalent bonding motifs always need counterions. Metavalent bonding has been classified in crystals that are composed of two elements with a minor electronegativity difference\cite{wuttigIncipientMetalsFunctional2018a}. Here the delocalized bonding extends over the whole crystal, as is reflected in our models. This analysis does not prove a difference between the two types; it merely shows that one language is used more dominantly in certain types of systems.

\section{Spectra, optical response, and sum rules}
\label{app:correlator_derivation}

\subsection{Energy-resolved tensor}
The quantum geometric tensor $\mathcal Q^{\mu\nu}=N_e^{-1}\operatorname{Tr}(P\mathcal D^\mu\mathcal D^\nu)$ collects ground-state transition-dipole correlations per electron, as defined in the main text. These correlations are present without thermal motion or an applied field. To resolve these correlations by excitation energy, we let one dipole operator evolve in time, $\mathcal D^\mu(t)=e^{iHt/\hbar}\mathcal D^\mu e^{-iHt/\hbar}$, and Fourier transform the correlation function $\mathcal Q^{\mu\nu}(t)=N_e^{-1}\operatorname{Tr}[P\mathcal D^\mu(t)\mathcal D^\nu(0)]$ to get
\begin{align}
\mathcal Q^{\mu\nu}(E)
&=\frac{1}{2\pi\hbar}\int_{-\infty}^{\infty}dt\,
e^{iEt/\hbar}\mathcal Q^{\mu\nu}(t)
=\frac{1}{N_e}
\sum_{\substack{n\in\mathrm{occ.}\\m\in\mathrm{unocc.}}}
r^\mu_{nm}r^\nu_{mn}\,\delta[E-(\epsilon_m-\epsilon_n)],
\label{eq:app_spectral_density}
\end{align}
with $r^\mu_{nm}=\langle\psi_n|\hat r^\mu|\psi_m\rangle$, so that $\int_0^\infty dE\,\mathcal Q^{\mu\nu}(E)=\mathcal Q^{\mu\nu}$. 
At zero temperature, the fluctuation--dissipation theorem identifies this positive-energy spectrum with the absorptive part of the electronic response, given in Section~\ref{app:ladder}.
The spectral tensor has units of length squared per energy. Inserting a delta function into Eq.~\eqref{eq:velocity_partition_si} defines the bond-resolved spectrum $Q^{\mu\nu}_{\alpha\beta\gamma\delta}(E)$. Its integral over energy is the bond term of Eq.~\eqref{eq:velocity_partition_si}, and its sum over orbitals is Eq.~\eqref{eq:app_spectral_density}.

\subsection{Optical sum rules}
\label{app:ladder}
We define the moments of the dipole spectrum as
\begin{equation}
M_p^{\mu\nu} = \int_0^\infty dE\,E^p\,\mathcal Q^{\mu\nu}(E)
=\frac{1}{N_e}\sum\limits_{\substack{n\in\mathrm{occ.}\\ m\in\mathrm{unocc.}}} (\epsilon_m-\epsilon_n)^p\, r^\mu_{nm}r^\nu_{mn},\qquad p=-1,0,1,
\label{eq:app_unified_moments}
\end{equation}
which follow from time-dependent QGT \cite{vermaInstantaneous2025}.
The moments keep the full complex tensor, including its antisymmetric part. 
The zeroth moment depends only on the occupied states. The other two also need the excitation energies, so they are not determined by the density matrix alone.

For a periodic insulator with electron density $n=N_e/V$, the induced polarization $\boldsymbol{\mathcal P}(\omega)=\epsilon_0\boldsymbol\chi_e(\omega)\mathbf E(\omega)$ gives the conductivity $\boldsymbol\sigma=-i\omega\epsilon_0\boldsymbol\chi_e$. 
The absorptive part of the conductivity at $E=\hbar\omega>0$ follows from the dipole spectrum, and the static electronic susceptibility follows via the Kramers--Kronig relation
\begin{align}
\sigma^{\mathrm{abs},\mu\nu}(\omega)
&=\frac{\pi e^2 n}{\hbar}E\mathcal Q^{\mu\nu}(E),
\label{eq:app_sigma_channel}\\
\chi_e^{\mathrm{abs},\mu\nu}(\omega)
&=\frac{\pi e^2n}{\epsilon_0}\mathcal Q^{\mu\nu}(E),
\qquad \boldsymbol\chi_e^{\mathrm{abs}}
=\frac{\boldsymbol\chi_e-\boldsymbol\chi_e^\dagger}{2i},
\label{eq:app_spectral_susceptibility}\\
\chi_{e,\mathrm{reg}}^{(\mu\nu)}(0)
&=\frac{2e^2n}{\epsilon_0}\operatorname{Re}M_{-1}^{\mu\nu}.
\label{eq:app_static_chi}
\end{align}
Here $\sigma^{\rm abs}$ is the absorptive, Hermitian part of the conductivity tensor. Its real symmetric part describes the absorption of linearly polarized light, and its imaginary antisymmetric part the difference in absorption between the two circular polarizations. 
The electronic dielectric constant is $\epsilon_r^{\mu\nu}=\delta_{\mu\nu}+\chi_e^{\mu\nu}$. For a molecule, replacing $n/\epsilon_0$ by $N_e$ in Eq.~\eqref{eq:app_static_chi} gives the static polarizability of second-order perturbation theory. 
These expressions describe independent electrons with the atoms held fixed, and omit excitonic and local-field effects.

The zeroth moment is the quantum metric, $g^{\mu\nu}=\operatorname{Re}M_0^{\mu\nu}$, and the first gives the total oscillator strength, or equivalently the inverse optical mass, $(m_{\rm opt}^{-1})^{\mu\nu}=2\operatorname{Re}M_1^{\mu\nu}/\hbar^2$.
The imaginary part of the first moment gives the self-rotation part of the orbital magnetic moment per electron, $\mu^z_{\rm SR}=(e/\hbar)\operatorname{Im}M_1^{xy}$, through the dichroic sum rule\cite{souzaDichroic$f$sumRule2008}. The full orbital magnetization contains an additional itinerant contribution\cite{ceresoliOrbitalMagnetization2006}.
In terms of the absorptive conductivity, the metric weights the absorption by $1/E$ and the static susceptibility by $1/E^2$.
In an orbital model with a finite number of bands, $M_1$ need not equal the free-electron value of the $f$-sum rule.

\subsection{Bond currents}
\label{app:bond_currents}
The charge flow across a bond follows from the change of the orbital populations. 
The population of orbital $\alpha$ is $\rho_{\alpha\alpha}=\operatorname{Tr}(\Pi_\alpha\rho)$
with $\Pi_\alpha=|\phi_\alpha\rangle\langle\phi_\alpha|$, and $\rho$ evolves as
$i\hbar\,\dot\rho=[H,\rho]$. Writing the change of $\rho_{\alpha\alpha}$ as a sum over the orbitals gives
\begin{equation}
\dot\rho_{\alpha\alpha}=\sum_\beta I^N_{\beta\to\alpha},
\qquad
I^N_{\beta\to\alpha}=\frac{2}{\hbar}\operatorname{Im}\!\left(H_{\alpha\beta}\rho_{\beta\alpha}\right),
\qquad
I_{\beta\to\alpha}=-e\,I^N_{\beta\to\alpha},
\label{eq:bond_current_mean_si}
\end{equation}
where $I^N_{\beta\to\alpha}$ is the number of electrons per unit time flowing from $\beta$ into $\alpha$ and $I_{\beta\to\alpha}$ the corresponding charge current, with $e>0$. 
Both are real and change sign
when the bond is reversed. A stationary state, $[H,\rho]=0$, therefore has zero net flow
into each orbital while individual bond flows can circulate around loops. 
With time-reversal symmetry and without spin--orbit coupling, $H$ and $\rho$ can be chosen real and every bond flow vanishes. When a magnetic field or magnetic order breaks time reversal, the bond flows can circulate, and around a molecular ring they are the ring currents of H\"uckel--London theory\cite{london1937,pople1958}.
The mean current requires the Hamiltonian as well as the density matrix, whereas the loop
products of Section~\ref{app:qgt_algebra} require only the density matrix.

Each bond flow is the expectation value of a Hermitian one-particle operator,
$I^N_{\beta\to\alpha}=\operatorname{Tr}(\rho\,\hat I^N_p)$, built from the hopping between the two orbitals,
\begin{equation}
\hat I^N_p=-\frac{i}{\hbar}\left(H_{\alpha\beta}\,|\phi_\alpha\rangle\langle\phi_\beta|
-H_{\beta\alpha}\,|\phi_\beta\rangle\langle\phi_\alpha|\right),
\qquad \hat I_p=-e\,\hat I^N_p,
\label{eq:bond_current_operator_si}
\end{equation}
where $p=(\beta\to\alpha)$ fixes one orientation for each unordered pair. These operators
are the bond pieces of the velocity. With $R_p^\mu=r_\alpha^\mu-r_\beta^\mu$, the
orbital-center approximation gives
\begin{equation}
v_{\alpha\beta}^\mu+v_{\beta\alpha}^\mu=R_p^\mu\,\hat I_p^N,
\qquad -e\,v^\mu=\sum_p R_p^\mu\,\hat I_p.
\label{eq:bond_current_velocity_si}
\end{equation}
The current operator $-e\,v^\mu$ is therefore the sum of the bond currents weighted by the corresponding bond vector.

\section{Loop contributions and symmetry}
\label{sec:chern-motifs-v15}
Here we decompose the Chern number of the Haldane model into contributions of individual sites and triangles. We use this decomposition to define the site contributions $c_a$ of Fig.~\ref{fig:loop_chemistry_summary}\textbf{A--C} and the family shares quoted in the main text, and to show that inversion partners add while same-sublattice triangles cancel.

\subsection{Relation to the Bianco--Resta local Chern marker} To relate the loop contributions to the local Chern marker introduced by Bianco and Resta\cite{biancoMappingTopologicalOrder2011} in a periodic crystal, we specify the unit cell containing each orbital and its embedding within that cell. We place atom $a$ in the home cell, $\mathbf R_a=0$, and denote by $\mathbf R_b$ and $\mathbf R_c$ the lattice vectors of the cells containing $b$ and $c$. The density-matrix block $\rho_{ab}(\mathbf R_b)=\langle a,0|P|b,\mathbf R_b\rangle$ depends only on the relative cell vector by translation symmetry.

A triangle anchored at $a$ then has vertices $\mathbf r_a=\boldsymbol\tau_a$, $\mathbf r_b=\mathbf R_b+\boldsymbol\tau_b$, and $\mathbf r_c=\mathbf R_c+\boldsymbol\tau_c$, where the $\boldsymbol\tau$'s are the atomic embeddings within the unit cell. Its three-center bond index and oriented area are
\begin{equation}
\begin{aligned}
b_{abc}(\mathbf R_b,\mathbf R_c)&=\operatorname{Tr}
[{\rho_{ab}(\mathbf R_b)\rho_{bc}(\mathbf R_c-\mathbf R_b)\rho_{ca}(-\mathbf R_c)}],\\
\mathcal A_{abc}^{xy}&=\tfrac12
[(\mathbf r_b-\mathbf r_a)\times(\mathbf r_c-\mathbf r_a)]_z.
\end{aligned}
\label{eq:motifs-product-area-v15}
\end{equation}
Here $a,b,c$ label physical atomic sites within the unit cell, with their periodic copies specified by the cell vectors. For an atomic block, $\operatorname{Tr}$ sums over its orbital and spin indices, so $b_{abc}(\mathbf R_b,\mathbf R_c)=\sum_{\alpha\in a,\beta\in b,\gamma\in c}b_{\alpha\beta\gamma}$.

 To assign the loops to a site $a$, we expand the projected positions in the site basis,
\begin{equation}
\begin{aligned}
\operatorname{Im}\operatorname{Tr}[(PxPyP)_{aa}]
&=\sum_{b,c}x_by_c\operatorname{Im}b_{abc}
=\sum_{b,c}\mathcal A_{abc}^{xy}\operatorname{Im}b_{abc}
=-\operatorname{Im}\operatorname{Tr}[(PxQyP)_{aa}],
\end{aligned}
\label{eq:bond_loop_local_marker_v15}
\end{equation}
where $b$ and $c$ run over all sites, including periodic images, and $Q=1-P$. 
Notice that reversing the traversal conjugates the loop product, $b_{acb}=b_{abc}^*$, so only the antisymmetric combination $(x_by_c-x_cy_b)/2$ contributes to the imaginary part.
The last equality holds because $\operatorname{Tr}[(PxyP)_{aa}]$ is real.

We write the Chern number as a sum of triangle weights, \begin{equation} w_{abc}=\frac{8\pi}{A_c}\mathcal A_{abc}^{xy}\operatorname{Im}b_{abc},\qquad C=\frac12\sum_{a\in\mathrm{cell}}\sum_{b,c\in\mathrm{lattice}}w_{abc}. \label{eq:motifs-weight-v15}\end{equation} Here $A_c$ is the cell area, $a$ runs over one cell, and $b,c$ run over the full lattice, including periodic copies; cell-vector arguments are suppressed. The factor $1/2$ removes double counting of the two traversals. Each triangle contributes once for each of its three vertices as anchor, translated into the home cell. 
With $C=2\pi n_e\Omega^{xy}$ (section~\ref{app:normalization}), the same sum can be written with the cell sum explicit,
\begin{equation}
C=-4\pi n_e\operatorname{Im}\mathcal Q^{xy}
=\frac{4\pi}{A_c}\sum_{a\in\mathrm{cell}}\operatorname{Im}\operatorname{Tr}[(PxPyP)_{aa}].
\label{eq:motifs-normalization-v15}
\end{equation}
Up to normalization, the site sum of Eq.~\eqref{eq:bond_loop_local_marker_v15} is the local marker of Bianco and Resta \cite{biancoMappingTopologicalOrder2011}, 
\begin{equation}
    -2\pi i\operatorname{Tr}([PxP,PyP]_{aa})=4\pi\operatorname{Im}\operatorname{Tr}[(PxPyP)_{aa}],
\end{equation}
which carries units of area. 
Dividing it by the sample area $N_{\mathrm{cell}}A_c$ gives the dimensionless site contribution $c_a$ of the main text. 
For a periodic sample of $N_{\mathrm{cell}}$ identical cells,
\begin{equation}
c_a=\frac{4\pi}{N_{\mathrm{cell}}A_c}\operatorname{Im}\operatorname{Tr}[(PxPyP)_{aa}]
=\frac{1}{2N_{\mathrm{cell}}}\sum_{b,c}w_{abc},
\label{eq:local_chern_contribution_v15}
\end{equation}
with $\sum_{a\in\mathrm{sample}}c_a=C$ and $\sum_{a\in\mathrm{one\ cell}}c_a=C/N_{\mathrm{cell}}$.
Each $c_a$ is thus a signed share of the Chern number. In an open flake or a molecule the same definition applies, and the complete sum vanishes (section~\ref{app:molecular}).
Triangles related by translation, rotation, inversion, or a change of anchor carry the same weight. At the parameters of Figure~4 the largest share, about $42\%$, comes from the family with two nearest-neighbor bonds and one second-neighbor contact (Fig.~\ref{fig:motifs-haldane-v15}\textbf{A}).

\subsection{Consequences of Inversion symmetry}
\label{sec:loops}
\begin{figure}
    \centering
    \includegraphics[width=0.5\linewidth]{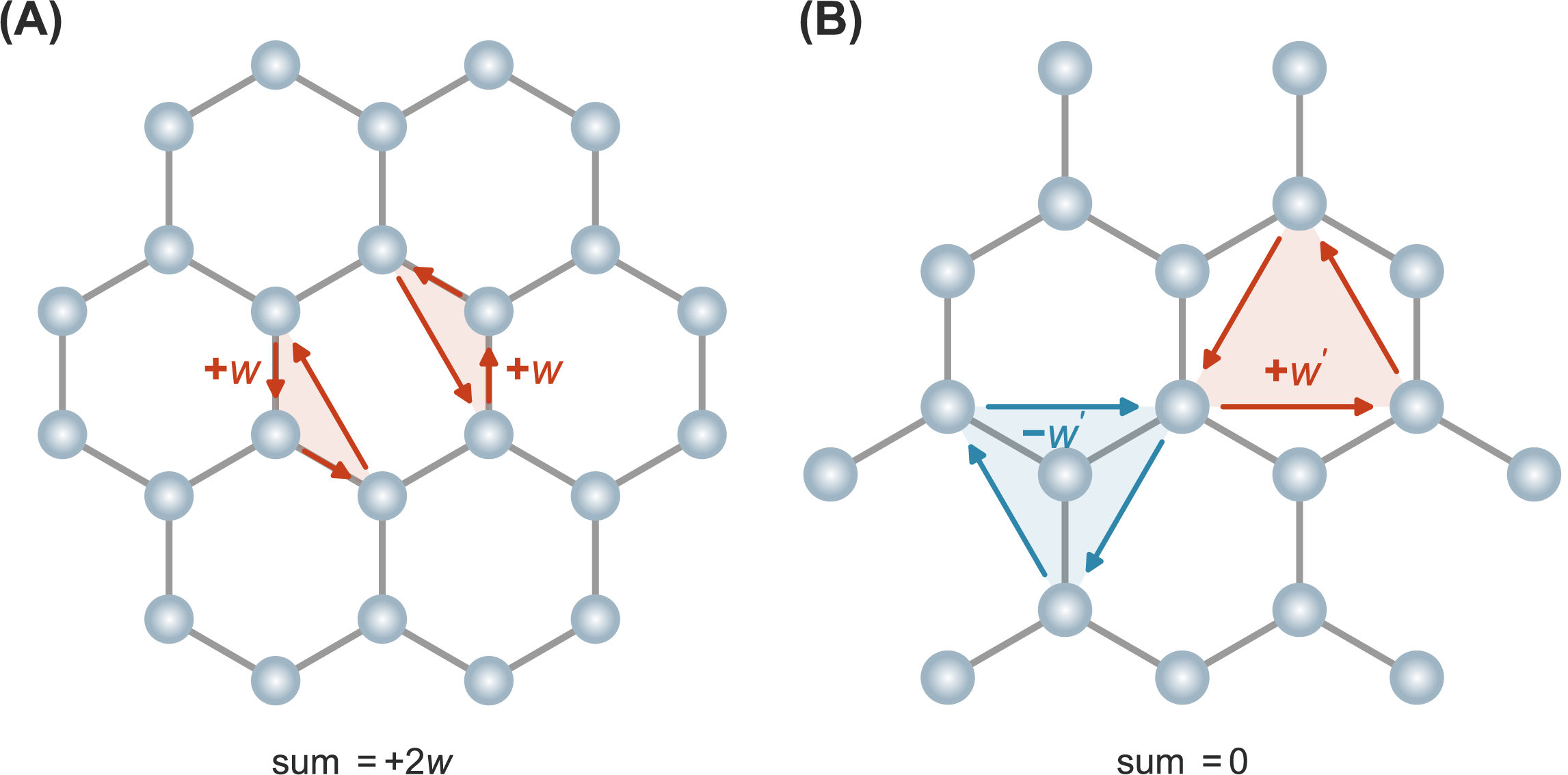}
    \caption{\textbf{Constructive and destructive addition of bonding loops in the Haldane lattice.} (\textbf{A}) The leading family of triangles, consisting of loops with two nearest-neighbor bonds and one second-neighbor contact joining the two sublattices, and its inversion partner. Both carry weight $+w$, so the pair adds. (\textbf{B}) Two same-sublattice triangles reflected through their shared site have equal areas and weights $\pm w'$ cancel. The inversion and fixed-anchor cancellation identities are given in section~\ref{sec:chern-motifs-v15}.}
\label{fig:motifs-haldane-v15}
\end{figure}

With equal energies on sublattices $A$ and $B$, the lattice is inversion symmetric. Inversion maps each triangle onto a partner with the same three-center bond index and the same oriented area, so the two contributions add (Fig.~\ref{fig:motifs-haldane-v15}\textbf{A}).
Same-sublattice triangles cancel for a different reason. For $\phi=\pi/2$ the nearest-neighbor hopping is real and the second-neighbor hopping purely imaginary, so that with $\Sigma=\operatorname{diag}(1,-1)$ on the two sublattices the Hamiltonian and, at gapped half filling, the projector obey
\begin{equation}
H^*=-\Sigma H\Sigma,\qquad P^*=1-\Sigma P\Sigma.
\label{eq:motifs-haldane-special-v15}
\end{equation}
Cross-sublattice elements of $P$ are therefore real, and off-diagonal same-sublattice elements purely imaginary. 
Together with Hermiticity and translation symmetry, this gives for $\mathbf R_b\neq0$
\begin{equation}
\rho_{AA}(-\mathbf R_b)=\rho_{AA}(\mathbf R_b)^*=-\rho_{AA}(\mathbf R_b).\end{equation}
Reflecting an AAA triangle through its anchor negates all three displacements, which flips the sign of its bond index but leaves its area unchanged. The two reflected triangles thus cancel, provided the distance cutoff contains both. 
For the alternating triangle of a hexagon, the reflected partner is the congruent triangle centered on an atom of the other sublattice, which the benzene graph lacks. 
Mixed-sublattice loops, however, have no reflected partner, since reflection through an A site does not map B sites onto atoms. With two real cross-sublattice links and one imaginary same-sublattice link, their bond indices are imaginary and survive the periodic sum.

\subsection{From ring to net: metric and fluctuations of the honeycomb models}
\label{sec:honeycomb_fluctuation}

\begin{figure}
    \includegraphics[width=1\linewidth]{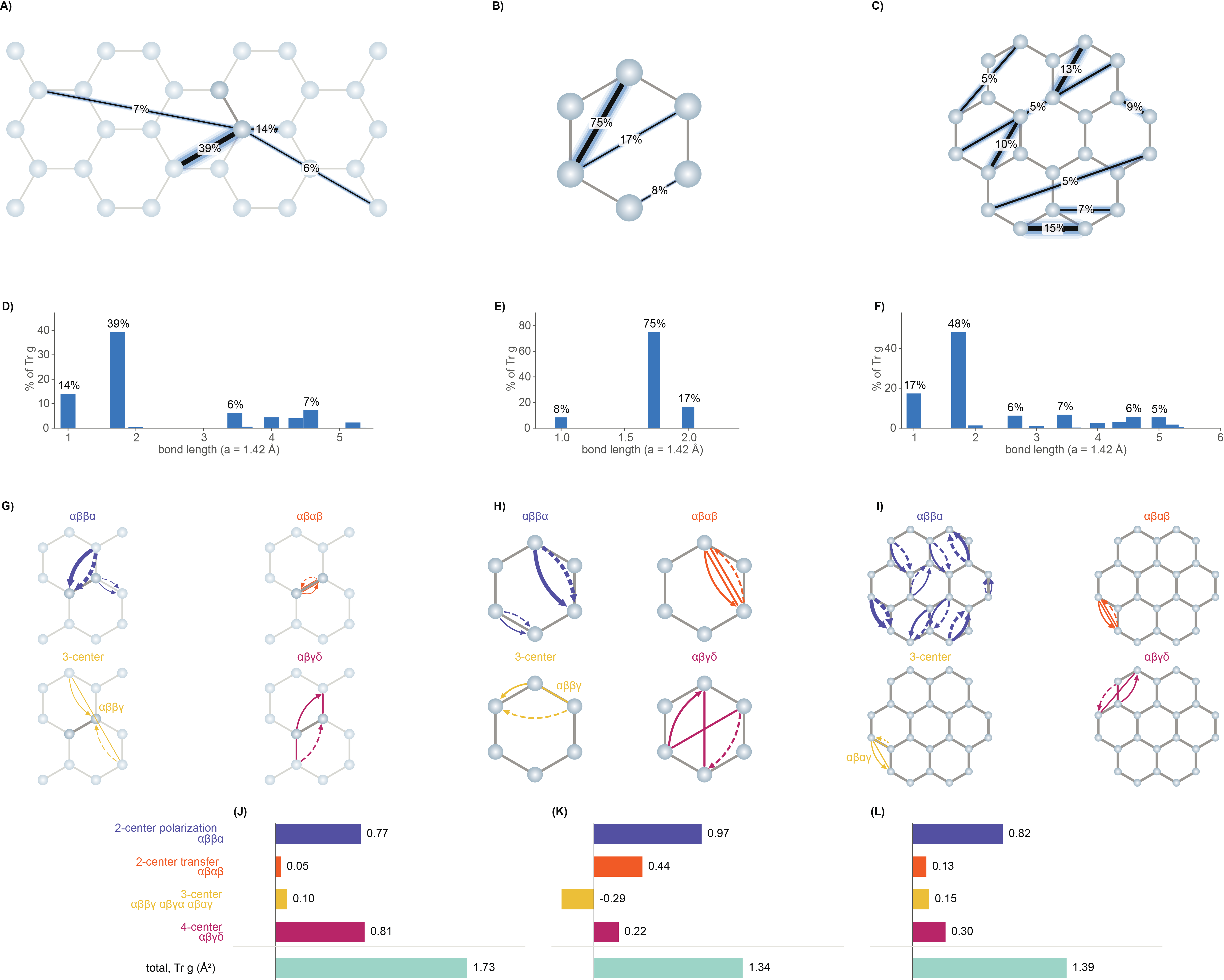}
    \caption{\textbf{Bond- and fluctuation-resolved quantum metric of the Haldane model, benzene and coronene.} (\textbf{A--C}) Bond-resolved quantum metric of (\textbf{A}) the Haldane model, (\textbf{B}) benzene and (\textbf{C}) coronene. Each bond type carrying $\ge5\%$ of $\operatorname{Tr}g$ is drawn once and labeled with the share of $\operatorname{Tr}g$ carried by it and its symmetry-equivalent bonds. The line width is proportional to this share. Drawn bonds account for 67\%, 100\% and 70\% of $\operatorname{Tr}g$ for the Haldane model, benzene and coronene. (\textbf{D--F}) Share of $\operatorname{Tr}g$ by bond length for (\textbf{D}) the Haldane model, (\textbf{E}) benzene and (\textbf{F}) coronene. (\textbf{G--I}) Fluctuation-resolved quantum metric for (\textbf{G}) the Haldane model, (\textbf{H}) benzene and (\textbf{I}) coronene. For each model, the largest fluctuations of each fluctuation class are drawn. The line width is linear in the contribution of all symmetry-equivalent fluctuations. (\textbf{J--L}) Breakdown of $\operatorname{Tr}g$ into fluctuation classes for (\textbf{J}) the Haldane model, (\textbf{K}) benzene and (\textbf{L}) coronene.}
     \label{fig:SI_benzene_fluctuation}
\end{figure}

In this section, we apply the bond and fluctuation decompositions to the three honeycomb $\pi$ systems of Fig.~\ref{fig:loop_chemistry_summary}: the periodic Haldane lattice, benzene and coronene. All use half filling, $t' = t$ and $\phi = \pi/2$. These are models with complex second-neighbor hopping, they should not be confused with zero-field molecules. With this analysis we can show what the metric records about electron sharing when only the geometry and the boundary conditions change.

The bond resolution [Fig.~\ref{fig:SI_benzene_fluctuation}{\textbf{A--F}}] shows that the metric does not follow the drawn bonds. In benzene the dominant contribution comes from the next-nearest-neighbor pairs instead of those where bonds are drawn. This is in qualitative agreement with the resonance model for benzene. If the bond picture were strictly followed, the metric would mimic a Kekul\'e structure, with which we mean the original description of benzene by Kekul\'e, of alternating single and double bonds arranged in a hexagon. Coronene keeps that same dominant shell but adds weight at every further separation. In the periodic lattice the leading shell holds only 14\% nearest neighbor 39\% next nearest neighbor contribution. The sequence ring, flake, net is a sequence of increasing delocalization, which can be seen directly in the metric tensor.

The fluctuation resolution [Fig.~\ref{fig:SI_benzene_fluctuation}{\textbf{G--L}}] shows how that delocalization is organized. Two-center polarization, $\alpha\beta\beta\alpha$, leads everywhere, since every bond polarizes. The difference lies in which multicenter pattern follows it. In benzene it is two-center transfer, $\alpha\beta\alpha\beta$, with the three-center term, $\alpha\beta\beta\gamma$ and its relatives, working against it: the ring closes on itself and the fluctuations have nowhere to go. In the periodic lattice this single-pair transfer nearly disappears and the four-center pattern, $\alpha\beta\gamma\delta$, becomes as strong as the polarization, meaning the ground state sustains coherence between two separate donor--acceptor pairs at once. Coronene lies between the two. Extended bonding is therefore not simply a longer-ranged version of molecular bonding; it changes which fluctuations dominate.

The total traces follow the same ordering, and together with the loop analysis of {section~\ref{sec:chern-motifs-v15}} they show that the same Hamiltonian gives a different quantum geometry once the boundary is removed.

\end{document}